\documentclass{aastex631} 
\submitjournal{ApJ}
\received{April 8, 2024}
\revised{May 30, 2024}
\accepted{Jun 4, 2024}
\usepackage[caption=false]{subfig}
\usepackage{multirow}
\usepackage{amsmath}
\let\longtable*\relax

\begin{document}

\title{Novel Atmospheric Dynamics Shape Inner Edge of Habitable Zone Around White Dwarfs}

\author{Ruizhi Zhan}

\affiliation{Dept. of Atmospheric and Oceanic Sciences, School of Physics, Peking University, Beijing 100871, People’s Republic of China}
\author{Daniel D.B. Koll}
\affiliation{Dept. of Atmospheric and Oceanic Sciences, School of Physics, Peking University, Beijing 100871, People’s Republic of China}
\correspondingauthor{Daniel D.B. Koll}
\email{dkoll@pku.edu.cn}
\author{Feng Ding}
\affiliation{Dept. of Atmospheric and Oceanic Sciences, School of Physics, Peking University, Beijing 100871, People’s Republic of China}

\begin{abstract}
White dwarfs offer a unique opportunity to search nearby stellar systems for signs of life, but the habitable zone around these stars is still poorly understood. Since white dwarfs are compact stars with low luminosity, any planets in their habitable zone should be tidally locked, like planets around M-dwarfs. Unlike planets around M-dwarfs, however, habitable white dwarf planets have to rotate very rapidly, with orbital periods ranging from hours to several days. Here we use the ExoCAM Global Climate Model (GCM) to investigate the inner edge of the habitable zone (HZ) around white dwarfs. Our simulations show habitable planets with ultrashort orbital periods ($P\lesssim$1 day) enter a ``bat rotation" regime, which differs from typical atmospheric circulation regimes around M dwarfs. Bat rotators feature mean equatorial subrotation and a displacement of the surface's hottest regions from the equator towards the midlatitudes. We qualitatively explain the onset of bat rotation using shallow water theory. The resulting circulation shifts increase dayside cloud cover and decrease stratospheric water vapor, expanding the white dwarf habitable zone by $\sim$50\% compared to estimates based on 1D models. The James Webb Space Telescope (JWST) should be able to quickly characterize bat rotators around nearby white dwarfs thanks to their distinct thermal phase curves. Our work underlines that tidally locked planets on ultrashort orbits may exhibit unique atmospheric dynamics, and guides future habitability studies of white dwarf systems.
\end{abstract}

\section{Introduction}

Searching for potential signs of life outside the Solar System is one of the main challenges of modern astronomy. By April 2024, there were more than 5600 confirmed exoplanets\footnote{https://exoplanets.nasa.gov}. Most of these planets orbit main sequence stars, such as Sun-like G-dwarfs or smaller M dwarfs. Some of them also orbit inside their star's habitable zone and have the right size to potentially sustain life. Nevertheless, the search for biosignatures remains inconclusive. This is largely because transmission spectroscopy, the most promising method for detecting biosignatures, strongly depends on the ratio between planet and stellar radius. Earth-sized planets are all much smaller than main sequence stars, which is why even optimistic estimates suggest that searching these planets for biosignatures will require major observational effort involving tens to hundreds of repeated transits \citep{morley_observing_2017,lustig-yaeger_detectability_2019,meadows_feasibility_2023}.

In this context, habitable planets around white dwarfs offer a unique opportunity \citep{agol_transit_2011, barnes_habitable_2013, cortes_detectability_2019, kaltenegger_white_2020,kozakis_high-resolution_2020}. White dwarfs are small, with radii approximately equal to Earth's, so planets orbiting these stars are much easier to characterize via transmission spectroscopy. Previous estimates suggest that, for a suitable white dwarf system, the James Webb Space Telescope (JWST) might be able to detect a biosignature gas like CH$_4$ in a single transit \citep{lin_h_2022}. Similarly, JWST should be able to detect stratospheric clouds on an Earth-sized planet inside a white dwarf's habitable zone within just four transits \citep{doshi_stratospheric_2022}.

To date there are no known habitable white dwarf planets. Nevertheless, multiple reasons suggest such planets could exist.
First, white dwarfs are common and make up about 5\% of nearby stars, which means they are about as abundant as Sun-like G dwarfs \citep{golovin_fifth_2023}.
Second, several non-habitable white dwarf planets have already been detected \citep{veras_planetary_2021}, including the transiting gas giant WD 1856+534 b \citep{vanderburg_giant_2020}. Third, although a star's habitable planets should be destroyed when the star leaves the main sequence and becomes a white dwarf, there are multiple ways to reform habitable planets. About half of all white dwarfs are polluted or have circumstellar discs \citep{koester_frequency_2014,putirka_polluted_2021,farihi_relentless_2022}, indicating many of them may be actively forming planets \citep{farihi_circumstellar_2016}. Distant planets, which survive the transition of their host star into a white dwarf, can also migrate inward \citep{veras_detectable_2015,munoz_kozai_2020,oconnor_enhanced_2020}, making it possible for these bodies to end up in the habitable zone.

Should white dwarfs indeed support habitable planets, they might end up providing the final refuge for life on galactic timescales. About 97\% of stars, including the Sun, will evolve into white dwarfs after the main sequence \citep{fontaine_potential_2001}. White dwarfs are born hot, with effective temperatures of $10^5$ to $10^4$ K, but cool rapidly via neutrino emission and thermal radiation. After cooling below 20000 K to 10000 K, white dwarfs stop emitting neutrinos and their cooling process slows down \citep{fontaine_potential_2001,bedard_spectral_2020}. Previous studies suggested that white dwarf systems can then remain habitable over timescales of $1-10$ Gyr, with a habitable zone that slowly shrinks as the star cools \citep{agol_transit_2011,becker_influence_2023}, which is opposite to their main sequence progenitors.

However, previous studies of white dwarf systems were based on idealized energy-balance models or 1D radiative-convective atmospheric models, and so did not account for atmospheric dynamics. The gold standard for habitability calculations are 3D Global Climate Models (GCMs), which self-consistently resolve the interaction between atmospheric dynamics, radiation, clouds, and ocean dynamics \citep{yang_stabilizing_2013,wolf_delayed_2014,hu_role_2014,    kopparapu_inner_2016,noda_circulation_2017,fujii_nir-driven_2017,komacek_atmospheric_2019,del_genio_habitable_2019,turbet_trappist-1_2021,yang_cloud_2023}. 
Habitable zone calculations of M dwarfs show that GCMs tend to predict significantly wider habitable zones than 1D models, which considerably increases the inferred frequency of habitable planets around main sequence stars \citep{yang_stabilizing_2013,wolf_delayed_2014,kopparapu_inner_2016,noda_circulation_2017,fujii_nir-driven_2017,komacek_atmospheric_2019,turbet_trappist-1_2021,yang_cloud_2023}.
In contrast, the impact of atmospheric dynamics on the white dwarf habitable zone is still unclear.

Like M dwarf planets, white dwarf planets should be tidally locked into synchronous rotation. Unlike M dwarfs, whose habitable zone corresponds to orbital periods of $\mathcal{O}(10)$ days, the habitable zone of white dwarfs corresponds to orbital periods as short as two hours. Only few GCM studies have considered tidally locked habitable planets with such rapid rotation rates, and these studies suggest that white dwarf planets might differ strongly from M dwarf planets \citep{merlis_atmospheric_2010,noda_circulation_2017,penn_atmospheric_2018,haqq-misra_demarcating_2018,komacek_atmospheric_2019,cohen_haze_2024}. For example, \citet{haqq-misra_demarcating_2018} showed that habitable M dwarf planets fall into three dynamical regimes: slow rotation (roughly, orbital periods of $P>20$ days), Rhines rotation ($5$ days $<P<20$ days), and rapid rotation ($P<5$ days). These regimes are associated with large differences in winds, heat transport, and cloud cover, modifying the M dwarf habitable zone. Similarly, \citet{noda_circulation_2017} investigated the impact of rotation on synchronously rotating planets and found that the atmospheric circulation changes strongly and abruptly at $P \sim 1$ days.

In this paper we therefore investigate the inner edge of the habitable zone around white dwarfs using a 3D GCM. We report a novel dynamical regime, the bat rotation regime, for planets with ultrashort orbital periods ($P \lesssim 1$ days) and investigate how it affects the inner edge of the habitable zone. Section \ref{method} describes the model and the assumed stellar and planetary parameters. Section \ref{bat} presents the bat rotation regime, compares it to other dynamical regimes of tidally locked planets, and uses shallow water theory to explain why the onset of bat rotation happens at $P \sim 1$ days. Section \ref{inneredge} shows that the bat rotation regime widens the habitable zone, by increasing dayside cloud cover and drying out the stratosphere; these effects push the Runaway Greenhouse Limit closer to the star and also suppress the Moist Greenhouse. Section \ref{discussion} expands beyond the inner edge of the habitable zone to show that most habitable planets around white dwarfs are likely to be bat rotators; it also discusses how our theoretical predictions can be tested using thermal phase curves with the James Webb Space Telescope (JWST). Finally, the paper concludes in Section \ref{conclusion}.

\section{Methods and Model}
\label{method}
\subsection{Stellar Configuration}
For the white dwarf host stars in our simulations, we used the most probable parameters based on Gaia EDR2 and EDR3 \citep{jimenez-esteban_white_2018,fusillo_catalogue_2021}. White dwarfs are assumed to be type DA (pure H envelope) with effective temperatures in the range 3500 $\leq$ $T_{\text{eff}}$ [K] $\leq$ 10000. For white dwarfs at 5000 K, 8000 K and 10000 K we use the default stellar spectra from ExoCAM\footnote{White dwarf spectra from ExoCAM: https://github.com/storyofthewolf/ExoRT/tree/main/data/solar}. 
ExoCAM does not provide white dwarf spectra at 3500 K and 6500 K, so for those stars we use spectral energy distributions (SEDs) based on 3D NLTE models from \cite{tremblay_spectroscopic_2013, tremblay_3d_2015}\footnote{White dwarf spectra from 3D NLTE model: https://warwick.ac.uk/fac/sci/physics/research/astro/people/tremblay/modelgrids}.
Note that the choice of stellar SED affects the modeled climates. Sensitivity tests show that, for the same stellar effective temperature, SEDs based on the 3D NLTE models from \citeauthor{tremblay_spectroscopic_2013} lead to somewhat higher surface temperatures than the default ExoCAM spectra (see Appendix \ref{appen:sensSED}).

All white dwarfs are assumed to have the same mass, $M = 0.6 M_\odot$, and radius, $a = 0.012 a_\odot $. These values are based on the mass-radius relationship $a_\text{WD}/a_{\odot} \approx 0.0127~(M_\text{WD}/M_{\odot})^{-1/3}~\sqrt{1-0.607 (M_\text{WD}/M_{\odot})^{4/3}}$ from \cite{veras_post-main-sequence_2016}. Here $a$ and $M$ are radius and mass, while subscript $\text{WD}$ and $\odot$ represent a white dwarf and the Sun. The surface gravity of a white dwarf with $0.6 M_\odot$ is $\log(g$[cm s$^{-2}$]$)= 8$.

\subsection{Planetary Configuration}

To compare our results with previous habitable-zone calculations for M dwarfs, we use the same model assumptions and parameters as in previous studies \citep[e.g.,][]{yang_strong_2014, kopparapu_inner_2016, kopparapu_habitable_2017}. We assume a planet with the same radius, mass, and surface gravity as Earth. For the planet's surface we assume an aquaplanet without land and a 50 m deep slab ocean (discussed in more detail in Section \ref{sect:3dgcm}). The atmosphere consists of only N$_2$ and H$_2$O. The orbital period $P$ is equal to
\begin{equation}
    \mathrm{P} = 365.25\ \mathrm{days}\ \left(\frac{M_\odot}{M_{\mathrm{WD}}}\right)^{\frac{1}{2}}\left( \frac{T_{\text{WD}}}{T_\odot}\right)^{3}\left( \frac{a_{\text{WD}}}{a_\odot}\right)^{\frac{3}{2}}\left( \frac{F_\oplus}{F_\mathrm{p}}\right)^{\frac{3}{4}},
    \label{orbitalperiod}
\end{equation}
which is derived from Kepler's third law. In Equation \ref{orbitalperiod}, $M$ is the stellar mass, $T$ is the stellar effective temperature, $a$ is the stellar radius, and $F$ is the flux received by the planet. The subscript $\odot$, $\oplus$, $\mathrm{WD}$, and $\mathrm{p}$ represent the Sun, Earth, white dwarf, and the planet orbiting the white dwarf, respectively.

In our simulations, $P$ lies in the range of 3 hours $\textless$ P $\textless$ 3.4 days. With such short $P$ and corresponding small semi-major axes, the timescales for tidal circularization and tidal locking around white dwarfs are less than 1000 years \citep{agol_transit_2011}. This is much shorter than the duration of the habitable zone, so we assume all planets are 1:1 tidally locked with zero obliquity and eccentricity. In addition we assume the planet is solely heated by the host star; other energy sources such as tidal and geothermal heating \citep{becker_influence_2023} are not included. Tidal heating requires non-zero orbital eccentricity or a non-tidally locked rotation state, so ignoring its effect is consistent with the assumed orbit and rotation state.

\subsection{3D Global Climate Model}
\label{sect:3dgcm}
The climate simulations use ExoCAM\footnote{ExoCAM source code: https://github.com/storyofthewolf/ExoCAM} \citep{wolf_exocam_2022}, a 3D GCM modified from the Community Atmosphere Model version 4 (CAM4) \citep{neale_description_2010}. ExoCAM uses ExoRT, a correlated-k radiative transfer package which employs the HITRAN 2016 database \citep{gordon_hitran2016_2017}. 

For the planet's surface we assume an aquaplanet without land and a 50 m deep slab ocean. The slab ocean is inert, so it stores heat locally but it does not redistribute heat horizontally. Previous studies indicated that ocean dynamics can have a significant impact on habitable planets, especially those near the outer edge of the habitable zone \citep{hu_role_2014,del_genio_habitable_2019,yang_ocean_2019}. However, here we focus on planets near the inner edge of the habitable zone. For such planets the planetary heat transport from day to nightside is dominated by the atmosphere \citep{yang_ocean_2019} and slab ocean simulations are sufficiently accurate to estimate important global-mean quantities, such as the water vapor mixing ratio in the upper atmosphere \citep{fujii_nir-driven_2017}.

In this work, ExoCAM is configured similar to the simulations in \cite{kopparapu_inner_2016}, \cite{komacek_atmospheric_2019}, and \cite{zhang_how_2020}. The initial condition is set to the ExoCAM default. The default spatial resolution is $4^\circ \times 5^\circ$ in latitude and longitude with 40 vertical levels from the surface up to 1 hPa. Based on sensitivity tests in which we changed the horizontal resolution (see Appendix \ref{appen:sensHOR}), the most rapidly rotating planets use a higher horizontal resolution of $1.9^\circ \times 2.5^\circ$; our model configurations are listed in Appendix \ref{appen:sensDET}. The reference albedo of the ocean surface at a zenith angle of 60$^\circ$ is 0.06 in the visible and 0.07 in the near-IR. The calculations use a time step of 30 minutes, with radiative transfer updated every two time steps.
All simulations are run to statistical equilibrium, which we define as when the global mean net top-of-atmosphere flux (absolute difference between incoming stellar and outgoing longwave) averaged over one Earth year reaches 3 W/m$^2$ or less. Typically, the GCM equilibrates within 35-45 Earth years. We analyze 10-Earth-year averaged outputs after statistical equilibrium.

\section{The Bat Rotation Regime}
\label{bat}

\subsection{New Dynamical Regime at High Rotation}

\begin{figure}[h!] 
    \centering
    \includegraphics[width=0.99\textwidth]{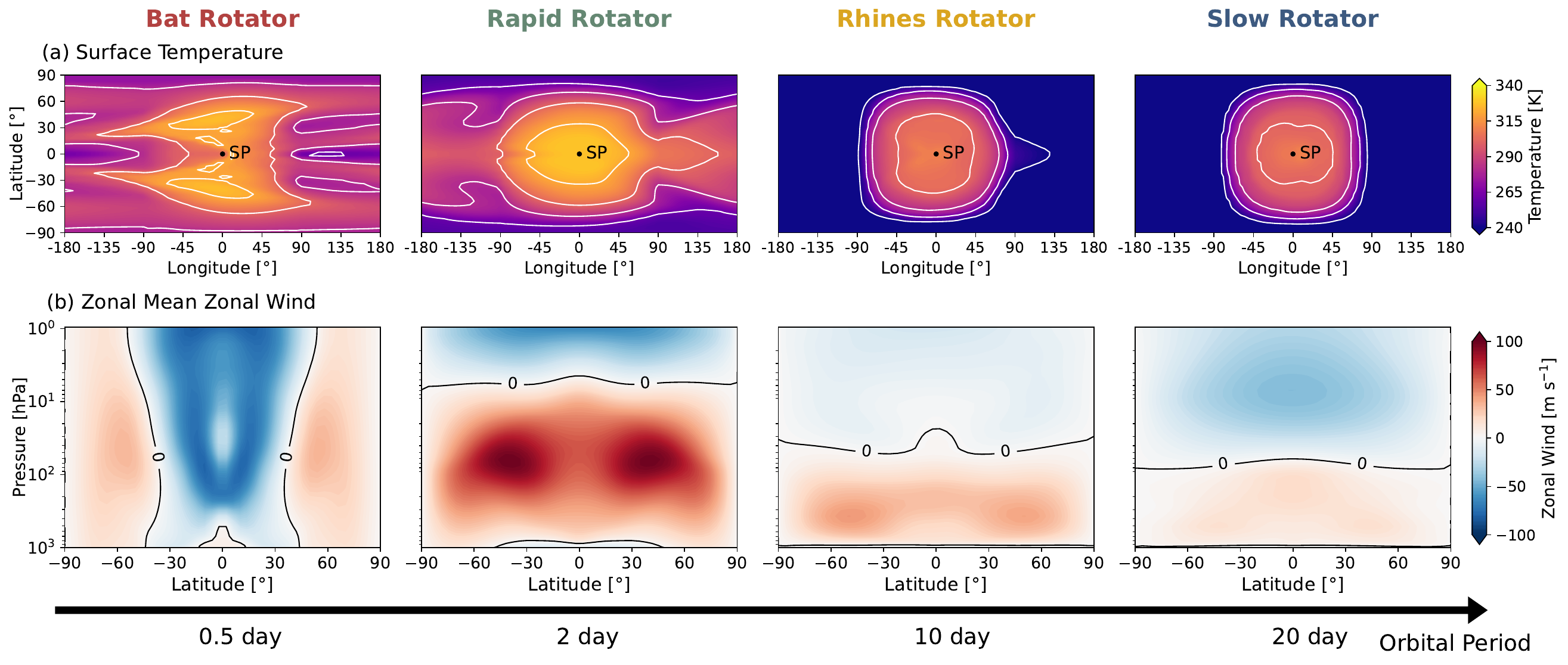}
    \caption{(a) Surface temperature and (b) zonal mean zonal wind as a function of rotation period. From left to right: bat rotator ($P=0.5$ days; this work), compared to a rapid rotator ($P=2$ days), Rhines rotator ($P=10$ days), and slow rotator ($P=20$ days). Dynamical regimes for the slower rotating planets are defined as in \citet{haqq-misra_demarcating_2018}. In (a), 'SP' corresponds to the substellar point. In (b), red versus blue shading represent eastward versus westward winds. The simulations assume a 10000 K white dwarf and 2.12 times Earth's instellation.}
    \label{batpattern-2}
\end{figure}

To explore the impact of rotation, we first fix the stellar spectrum and stellar constant in ExoCAM to that of a young white dwarf and only vary the planet's rotation rate. We consider rotation rates ranging from a slow 20 days, representative of a habitable planet around an early M dwarf, over a moderate 10 and 2 days, representative of a habitable planet around a late M dwarf, to a rapid 0.5 days, representative of a habitable planet around a white dwarf (sensitivity tests suggest that the model's horizontal resolution is sufficient to resolve the Rossby deformation radius even at such rapid rotation; see Appendix \ref{appen:sensHOR}).
Figure \ref{batpattern-2} shows the resulting surface temperature and zonal mean zonal wind maps.

Our simulations at slow and moderate rotation match those discussed in previous studies of M dwarf planets \citep{noda_circulation_2017,haqq-misra_demarcating_2018,komacek_atmospheric_2019,cohen_haze_2024}. Due to the close overlap between our results and those in \citet{haqq-misra_demarcating_2018} we use the same dynamical regime labels, namely slow rotator, Rhines rotator, and rapid rotator. Figure \ref{batpattern-2} shows that as rotation rate increases, surface temperature changes from being dominated by a hemispherical day-night temperature gradient (slow rotator, Rhines rotator), towards a zonally banded structure (rapid rotator). In all cases, the hottest point on the surface is at the equator. Similarly, the atmospheric circulation changes from a global day-night overturning cell with weak equatorial superrotation (slow rotator), towards a zonally dominated circulation with increased equatorial superrotation and a pair of strong eastward midlatitude jets (Rhines rotator, rapid rotator).

Surprisingly, as rotation increases further and exceeds one rotation per day, a new dynamical regime emerges (Figure \ref{batpattern-2}, left column). The hottest surface temperatures move from the equator to the mid-latitudes and take on the form of slanted zonal bands, while a cold tongue develops in the equatorial night hemisphere. Meanwhile, zonal mean zonal winds also change dramatically. Equatorial superrotation largely disappears, except in a region of remnant weak superrotation below 500 hPa. Above 500 hPa at the equator, as well as throughout the entire atmospheric column off the equator, the zonal winds reverse sign and the atmosphere develops a strong westward, i.e., subrotating, jet.
The fact that the hottest surface temperatures move off the equator and the zonal winds flip sign at the equator distinguish this simulation from the dynamical regimes of M dwarf planets discussed in \citet{haqq-misra_demarcating_2018}, but are consistent with the most rapidly rotating simulations ($P\lesssim$ 1 day) presented in previous theoretical studies of tidally locked planets \citep{merlis_atmospheric_2010,noda_circulation_2017,komacek_atmospheric_2019,cohen_haze_2024}.

\begin{figure}[h!] 
    \centering
    \includegraphics[width=0.65\textwidth]{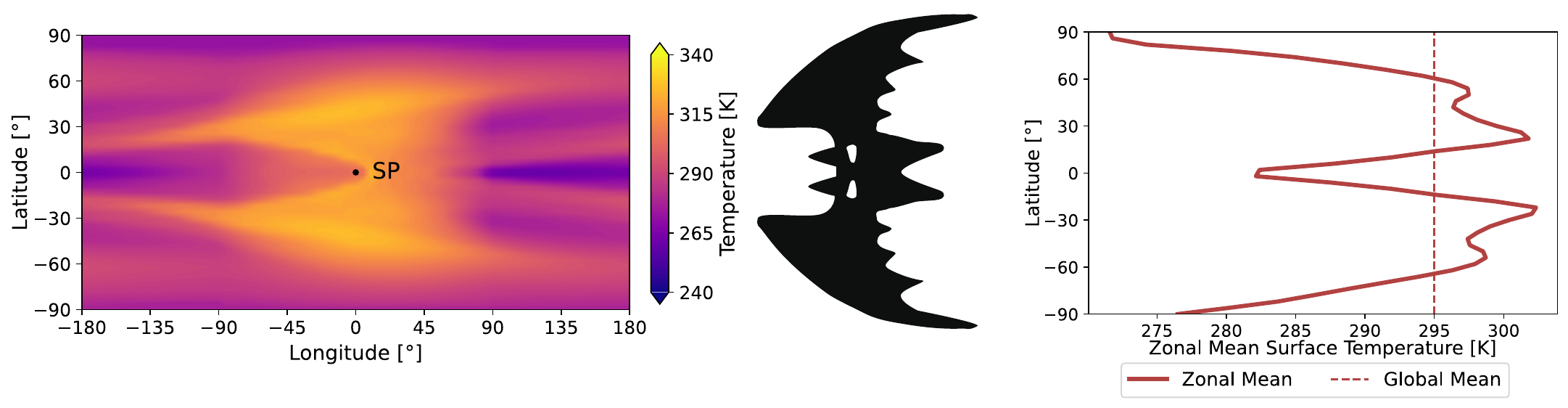}
    \caption{The surface temperature map and zonal mean surface temperature profile of a typical bat rotator. We name the bat rotation regime after the shape of its surface temperature map. Left: 'SP' shows the substellar point. Simulation shown is identical to Figure \ref{batpattern-2}, left column.}
    \label{batpattern-1}
\end{figure}

Based on its distinctive surface temperature pattern we name the most rapidly rotating simulation a `bat' rotator, as illustrated in Figure \ref{batpattern-1}. In the rest of this paper, we define bat rotators as planets with their hottest surface temperature located off the equator (Fig.~\ref{batpattern-1}, right panel) and with net subrotating winds at the equator. In practice we evaluate these criteria based on whether the latitude of the maximum in zonal mean surface temperature is more than one equatorial Rossby deformation radius away from the equator, and whether the (mass-weighted) vertical and zonal mean zonal wind within one Rossby deformation radius of the equator is westward,
\begin{subequations}
\begin{align}
    \left| y_{\mathrm{max}\ \bar{T}_s(\mathrm{lat})} \right| >\mathrm{L}_{\mathrm{Ro}},  \label{batdef} \\
    \bar{U}_{eq} < 0.
\end{align}
\label{batdef-all}
\end{subequations}

\noindent Here $y_{\mathrm{max}\ \bar{T}_s(\mathrm{lat})}$ is the distance between the latitude of the zonal mean surface temperature maximum and the equator, $\mathrm{L}_{\mathrm{Ro}}$ is the equatorial Rossby deformation radius, and $\bar{U}_{eq}$ is the vertical and zonal mean zonal wind near the equator. We evaluate $\mathrm{L}_{\mathrm{Ro}}$ as $\mathrm{L}_{\mathrm{Ro}} = \left( \frac{a}{2\Omega}\right)^{\frac{1}{2}}\left( \frac{RgH}{c_p} \right)^{\frac{1}{4}}$, which assumes an isothermal atmosphere; $a$ is the planet radius, $\Omega$ the planetary rotation rate, $R$ the specific gas constant, $g$ surface gravity, $H$ is the scale height, and $c_p$ is the specific heat of air at constant pressure.

\subsection{Features of the Bat Rotation Regime}
\begin{figure}[htbp]
    \centering
    \includegraphics[width=1.0\textwidth]{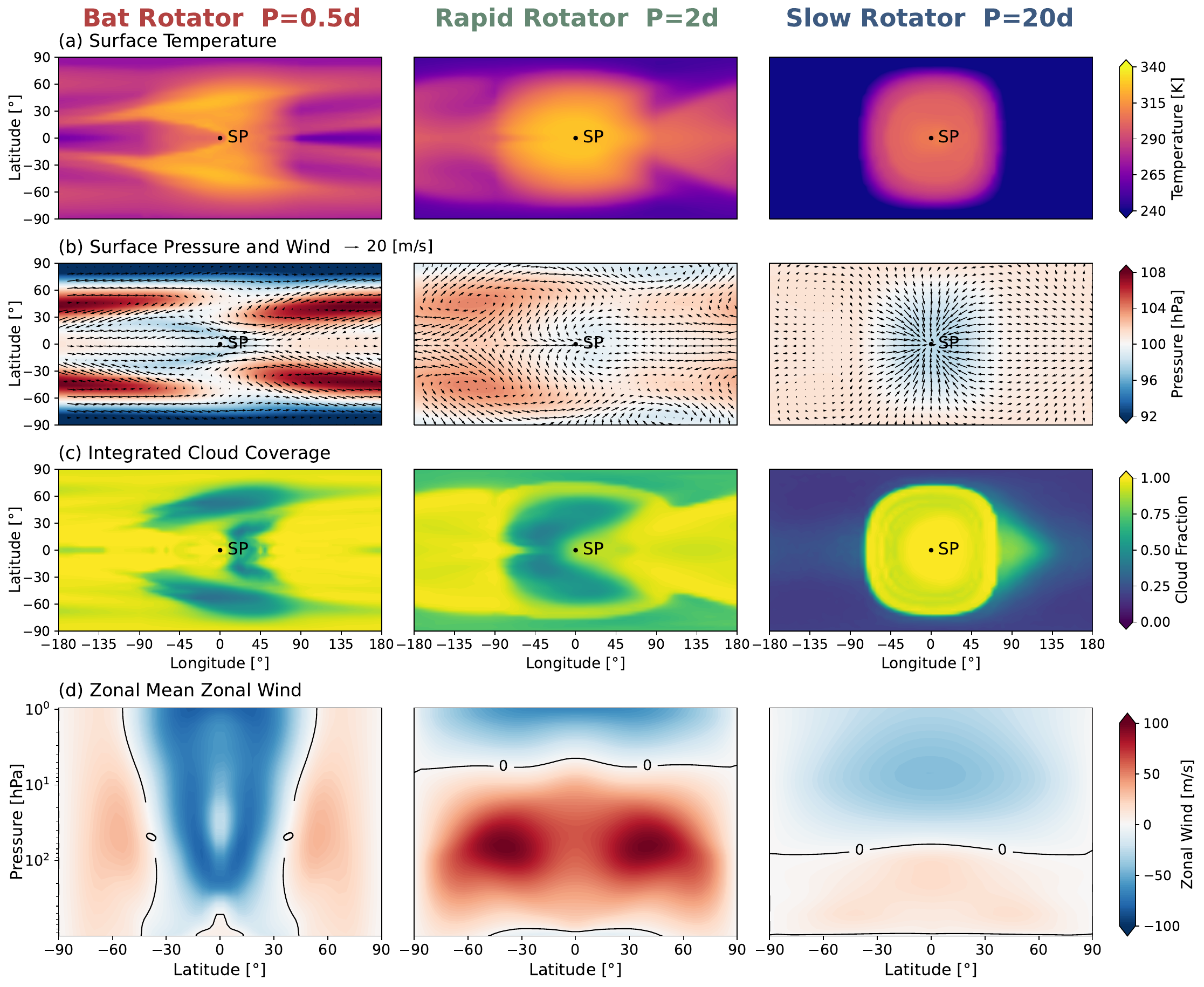}
    \caption{Climate comparison of a bat rotator, a rapid rotator and a slow rotator. 'SP' corresponds to the substellar point in each map. The three rotators are orbiting a 10000 K white dwarf and receiving 2.12 times Earth's insolation. Except for the orbital periods set at 0.5, 1.96, and 20 days respectively, the simulations are configured identically. From top to bottom panels are maps of surface temperature, surface pressure with surface wind at 970 hPa, total cloud fraction and zonal mean zonal wind.}
    \label{batandrapid}
\end{figure}

The emergence of a new dynamical regime at rapid rotation has broad implications for climates on white dwarf planets. Figure \ref{batandrapid} shows additional output variables from the idealized simulations in Figure \ref{batpattern-2}. To highlight the dominant impact of rotation, we do not include the Rhines rotator and only compare the bat rotator against rapid and slow rotators.

Figure \ref{batandrapid}b shows maps of surface pressure together with arrows indicating near-surface horizontal winds. On the slow rotator, surface pressure and surface temperature coincide closely. In line with a day-night thermally direct circulation, surface horizontal winds converge at the substellar point (SP) where surface pressure is lowest. On the rapid rotator a high pressure center forms on the nightside mid-latitudes, indicative of a stationary Rossby wave with zonal wavenumber one. Near-surface winds no longer directly point from high to low pressure, suggesting horizontal gradients are balanced by both friction and Coriolis forces (frictional-geostrophic balance). On the bat rotator, surface pressure differences are larger and mid-latitude winds are more closely aligned with pressure contours than on the rapid rotator, consistent with a flow dominated by purely geostrophic balance.

The surface pressure map of the bat rotator is consistent with its distinctively banded surface temperature pattern (compare Fig.~\ref{batandrapid}a and b). The high-pressure centers in the mid-latitudes are flanked by low-pressure centers at the poles. The resulting geostrophic flow around these centers transports sensible and latent heat to the night side which helps support the banded, bat-like, surface temperature pattern. Similarly, the pressure centers at low latitudes are twisted and banded, with a high pressure band on the nightside equator. Air sinks in this region and horizontal winds are weak, suggesting less exchange with heat from the dayside which helps to sustain the cold tongue at the equator.

Figure \ref{batandrapid}c shows maps of cloud fraction. Previous work found that more rapid rotation on M-dwarf planets weakens the convergence of zonal winds near the substellar point and decreases dayside cloud cover \citep{kopparapu_inner_2016,haqq-misra_demarcating_2018,komacek_atmospheric_2019}. Surprisingly, we find that the bat rotator is cloudier than the rapid rotator inside the substellar region (defined as within 30$^\circ$ of the substellar point; see Figure \ref{batandrapid}c). Consequently, the substellar region on a bat rotator receives less shortwave radiation, promoting cooler surface temperatures at the equator. In addition, there are banded cloud-free regions in the `bat wings' on the dayside between 30$^\circ$-60$^\circ$. These regions receive more shortwave radiation, consistent with the bat wings also containing the hottest surface temperatures (compare Fig.~\ref{batandrapid}a and c). 
Finally, consistent with equatorial superrotation on the rapid rotator versus subrotation on the bat rotator, the clouds near the substellar point are shifted east on the rapid rotator versus west on the rapid rotator (Fig.~\ref{batandrapid}b).

Figure \ref{batandrapid}d displays the zonal mean winds. The rapid rotator is dominated by two tropospheric eastward jets that peak in the mid-latitudes (30-60$^{\circ}$), and has superrotation at the equator. In contrast, the bat rotator is dominated by a double-peaked westward jet, with subrotation equatorwards of 45$^\circ$. Only near the surface, below 500 hPa, does the bat rotator still show weak equatorial superrotation. At high latitudes (45-90$^{\circ}$), the bat rotator also has two eastward jets, but the jets are moved polewards and are weaker than on the rapid rotator.

Overall, the dynamics of the bat rotation regime are thus quite distinct from the dynamics reported in most previous studies of tidally locked planets. In agreement with Rhines scale arguments which predict that jet width decreases with increasing rotation \citep[e.g.,][]{showman_atmospheric_2013}, the bat rotator features more zonal jets than the rapid rotator.
However, previous studies found that more rapid rotation strengthens equatorial superrotation \citep{merlis_atmospheric_2010,kopparapu_inner_2016,haqq-misra_demarcating_2018}, whereas here equatorial superrotation disappears almost entirely and is replaced by strong subrotation. The emergence of subrotation is consistent with thermal wind balance. On a bat rotator the mid-latitudes are hotter than the equator, so any geostrophically balanced low-latitude jet should flip from eastward to westward (note this argument does not hold directly at the equator where the Coriolis force is zero, consistent with the fact that the subrotating jet on the bat rotator is strongest off the equator; see Figure \ref{batandrapid}d).
Finally, the equator-to-pole temperature contrast on the bat rotator is also lower than on the rapid rotator (see Figure \ref{batandrapid}). This decreasing trend is opposite to that found in previous studies of both tidally locked and non-tidally locked planets \citep{kaspi_atmospheric_2015, haqq-misra_demarcating_2018}, which showed that higher rotation leads to smaller eddies and less efficient poleward energy transport.

\subsection{Distribution of Bat Rotators around White Dwarfs}

\begin{figure}[t]
    \centering
    \includegraphics[width=0.7\textwidth]{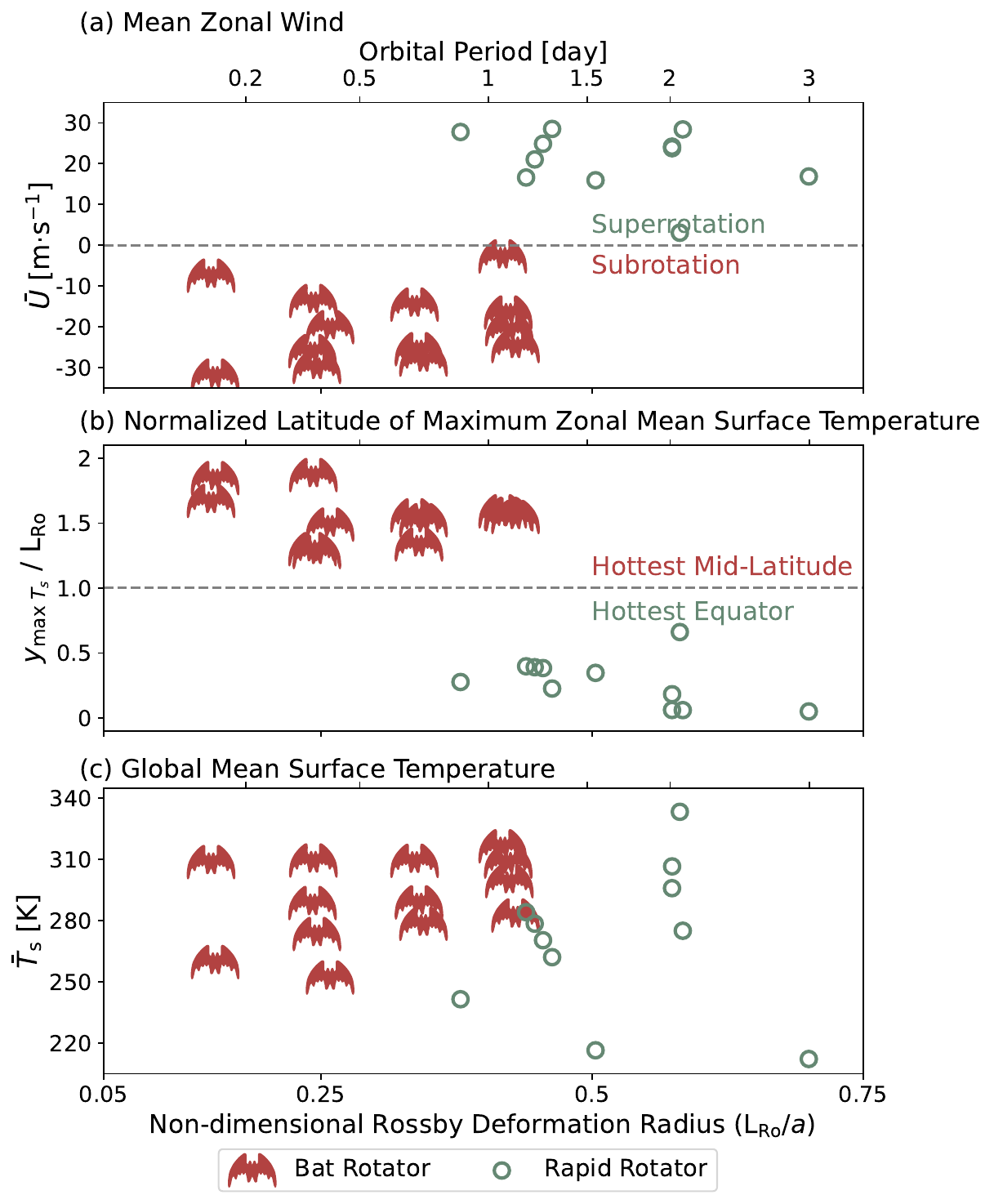}
    \caption{Distribution of bat rotators and rapid rotators in our simulations as a function of orbital period and non-dimensional Rossby deformation radius. (a) Zonal and vertical mean zonal wind near the equator (within one Rossby deformation radius from the equator). The vertical mean is computed using a mass-weighted average. (b) Distance between the equator and zonal mean surface temperature maximum, normalized by the equatorial Rossby deformation radius. (c) Global mean surface temperature. Red bat patterns and green circles represent bat rotators and rapid rotators, respectively.}
    \label{batdistribution}
\end{figure}

How significant is the bat rotation regime for white dwarf planets? To address this question, we explored different combinations of instellation and stellar effective temperature with the GCM. The assumed host stars range from hot young white dwarfs (10000 K) to cool old white dwarfs (3500 K). Planetary rotation rate is varied self-consistently, with orbital periods ranging from 3.4 days (coldest simulation orbiting a 10000 K white dwarf) down to 0.13 days (hottest simulation orbiting a 3500 K white dwarf); see Appendix Table \ref{table:configuration}.

Figure \ref{batdistribution} shows the resulting simulations as a function of rotation rate. The rotation rate is also expressed using the non-dimensional Rossby deformation radius, defined as the equatorial Rossby deformation radius $\mathrm{L_{Ro}}$ divided by the planetary radius $a$. In all cases the equatorial Rossby deformation radius is smaller than the planetary radius, $\mathrm{L_{Ro}}/a < 1$, which means equatorial waves can't propagate globally due to the influence of rotation \citep{leconte_3d_2013,koll_deciphering_2015}. To categorize the simulations, we classify a simulation as a bat rotator if it satisfies Equations \ref{batdef-all}; all other simulations are classified as rapid rotators.

We find little-to-no overlap between bat rotators and rapid rotators, supporting our classification of them as distinct dynamical regimes \citep[also see][]{noda_circulation_2017}. Figure \ref{batdistribution}a shows the equatorial zonal and vertical mean zonal wind $\bar{U}_{eq}$, defined above. All simulations either display strong equatorial superrotation or weak-to-strong subrotation; notably, we find no simulations with intermediate weak superrotation. Similarly, Figure \ref{batdistribution}b shows that the zonal-mean surface temperature maximum in all simulations is either located close to the equator, 
or significantly off the equator ($y_{\max Ts}/\mathrm{L}_{\mathrm{Ro}} = 1.5$ at $P=0.5$ days translates to about 26 degrees), with no intermediate values.
Finally, bat rotators tend to be hotter than rapid rotators (see Figure \ref{batdistribution}c). This is because instellation and orbital period are varied self-consistently, so planets that receive more stellar flux tend to be closer to their host stars and rotate more rapidly. Therefore, the bat rotation regime is most relevant for the inner edge of the habitable zone around white dwarfs.

\subsection{Predicting the Onset of Bat Rotation}

Figures \ref{batandrapid} and \ref{batdistribution} show that the transition from rapid rotators to bat rotators is abrupt and occurs at a critical rotation period of about 1 day. Can this critical rotation period be understood using theory? Similarly, why do our rocky planet simulations at $P\sim 1$ day develop equatorial subrotation and mid-latitude temperature maxima, whereas hot Jupiters at $P\sim 1$ day have equatorial superrotation and equatorial hot spots \citep{showman_equatorial_2011}? In this section we offer a tentative explanation based on shallow water theory.

Briefly, previous work developed analytic solutions to the shallow water equations on the equatorial beta plane \citep{gill_simple_1980,wu_thermally_2001,showman_equatorial_2011}. Below we use the linear solutions from \cite{showman_equatorial_2011} which consider the response of an atmospheric layer to heating and cooling, parameterized as Newtonian relaxation of the layer's thickness over a radiative timescale $\tau_{\mathrm{rad}}$. Friction is included as Rayleigh drag with timescale $\tau_{\mathrm{drag}}$, representing friction from the planetary surface, vertical turbulent mixing, or momentum transport by breaking gravity waves. The shallow water equations are nondimensionalized by the equatorial Rossby deformation radius ($L = \sqrt{\frac{\sqrt{gH}}{\beta}}$), the gravity wave phase speed ($U_\mathrm{wave} = \sqrt{gH}$) and the time for gravity waves to cross a deformation radius ($\mathcal{T} =\sqrt{\frac{1}{\sqrt{gH}\beta}}$). The linearized and nondimensionalized equations read
\begin{subequations}
\begin{align}
    \frac{\partial \eta}{\partial x}-yv=-\frac{u}{\hat{\tau}_{\mathrm{drag}}}, \label{swe-1} \\
    \frac{\partial \eta}{\partial y}+yu=-\frac{v}{\hat{\tau}_{\mathrm{drag}}}, \label{swe-2}\\
    \frac{\partial u}{\partial x}+\frac{\partial v}{\partial y}=S(x,y)-\frac{\eta}{\hat{\tau}_\mathrm{rad}}. \label{swe-3}
\end{align}
\label{swe-all}
\end{subequations}
Here $x$ and $y$ are eastward and northward distance, $u$ and $v$ are the eastward and northward horizontal velocity, $\hat{\tau}_{\mathrm{drag}}$ and $\hat{\tau}_{\mathrm{rad}}$ are dimensionless timescales, i.e. $\hat{\tau}=\frac{\tau}{\mathcal{T}}$, and $\eta$ is the deviation of the atmospheric layer's thickness from its reference thickness $H$. Note that if all other parameters are held fixed, increasing the planet's rotation (larger $\beta$) decreases $L$ and $\mathcal{T}$, which increases $\hat{\tau}_{\mathrm{drag}}$ and $\hat{\tau}_{\mathrm{rad}}$.

\begin{figure}[t]
    \centering
    \includegraphics[width=1.0\textwidth]{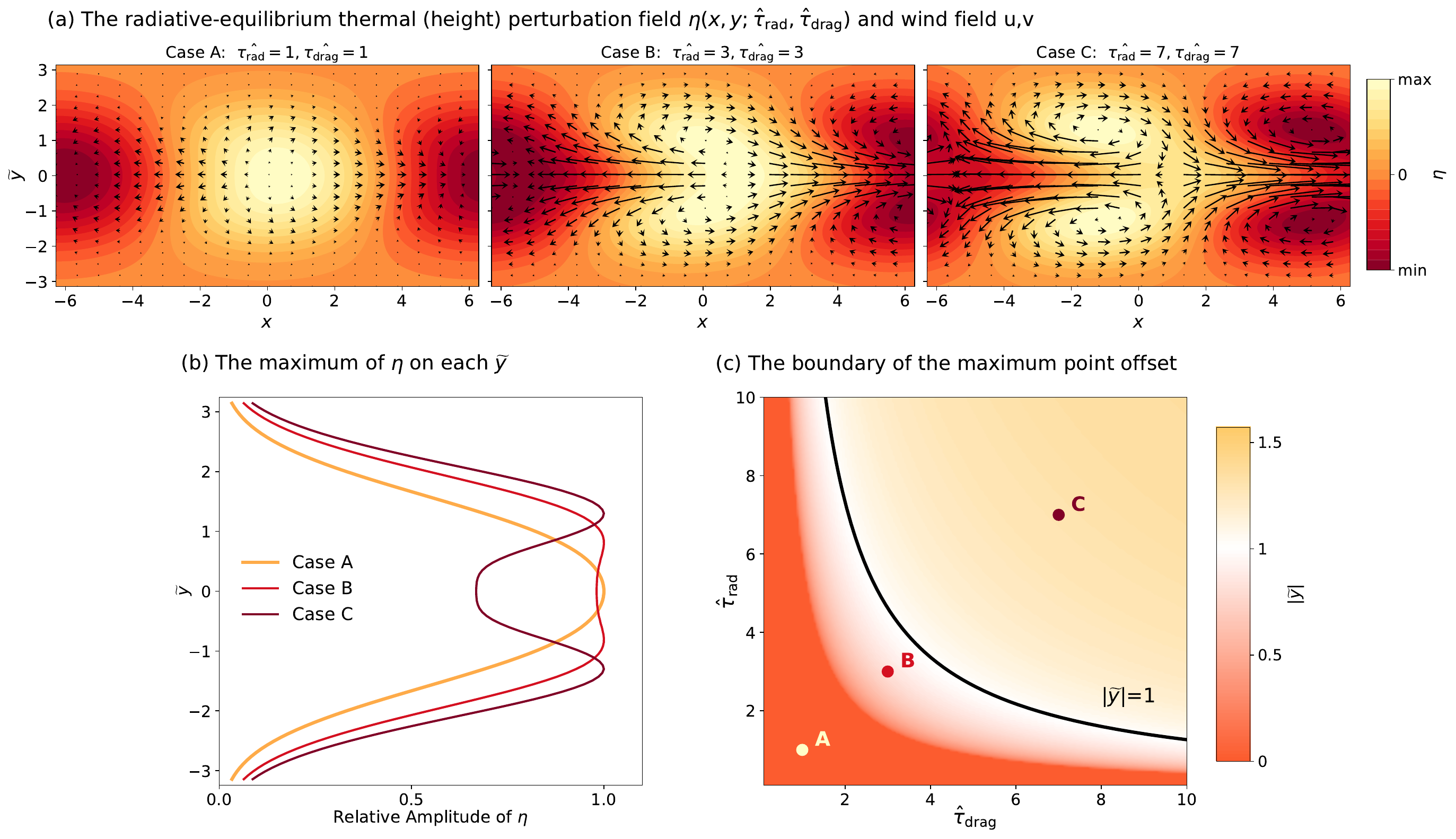}
    \caption{The shallow water equations qualitatively reproduce the transition from rapid to bat rotator. (a) Analytic solutions of the linearized shallow water equations (Eqns.~\ref{swe-1}-\ref{swe-3}) for dimensionless zonal wavenumber $k = 0.5$. Colors represent the height perturbation field, arrows represent horizontal winds. Assuming all other parameters are held fixed, rotation increases from Case A to C. (b) The maximum of $\eta$ at each non-dimensional latitude $\widetilde{y}$. The value of $\eta$ at each latitude is normalized by the global maximum of $\eta$. (c) Absolute value of the latitude at which $\eta$ is largest, $|\widetilde{y}|$, as a function of the dimensionless timescales $\hat{\tau}_{\mathrm{rad}}$ and $\hat{\tau}_{\mathrm{drag}}$. Black line shows $|\widetilde{y}|=1$. Dark colors correspond to the height maximum at the equator, light colors correspond to the height maximum at midlatitudes.}
    \label{SWE}
\end{figure}

For finite $\tau_{\mathrm{drag}}$, solutions to these equations take on the meridional structure of the parabolic cylinder functions $\psi_n(y)$ \citep{wu_thermally_2001,showman_equatorial_2011}
\begin{equation}
    \psi_n(y)=\exp{\left[ -\frac{1}{2}\left( \frac{y}{\mathcal{P}} \right)^2 \right]}H_n\left( \frac{y}{\mathcal{P}} \right),
\end{equation}
which are Gaussians times Hermite polynomials $H_n$; the first few Hermite polynomials are $H_0(x)=1, H_1(x)=2x, H_2(x)=4x^2-2$, and Hermite polynomials are related via $H_{n+1}(x)=2xH_n(x)-H'_n(x)$. Here $\mathcal{P}=\left( \frac{{\tau}_{\mathrm{rad}}}{{\tau}_{\mathrm{drag}}} \right)^{\frac{1}{4}}$ is the fourth root of a Prandtl number. For tidally locked planets, if the thermal forcing and solutions vary sinusoidally in longitude with dimensionless zonal wavenumber $k$, the pattern of the thermal forcing can be represented as 
\begin{equation}
    S(x,y)=\sum_{n=0}^{\infty}S_ne^{ikx}\psi_n(y),
\end{equation}
while the layer's thickness (proportional to the atmosphere's temperature) is represented as
\begin{equation}
    \eta(x,y)=\sum_{n=0}^{\infty}\hat\eta_n(y,\hat{\tau}_{\mathrm{rad}},\hat{\tau}_{\mathrm{drag}})e^{ -\frac{1}{2}\left( \frac{y}{\mathcal{P}} \right)^2}e^{ikx},
    \label{swe_eta}
\end{equation}
where $\hat\eta_n(\frac{y}{\mathcal{P}})$ are complex functions of $\frac{y}{\mathcal{P}}$.
Equation \ref{swe_eta} shows that $\eta(x,y)$ is determined by only two nondimensional parameters, $\hat{\tau}_{\mathrm{rad}}$ and $\hat{\tau}_{\mathrm{drag}}$. These are equivalent to three dimensional parameters, ${\tau}_{\mathrm{rad}}$, ${\tau}_{\mathrm{drag}}$, and $\mathcal{T}$, which represent radiative damping, frictional damping, and geostrophic adjustment.

To investigate the transition between rapid rotators and bat rotators, we evaluate solutions of $\eta(x,y)$ for different $\hat{\tau}_{\mathrm{rad}}$ and $\hat{\tau}_{\mathrm{drag}}$ following the methods in \citet{showman_equatorial_2011}. Figure \ref{SWE}a shows that, when the dimensionless timescales are short (at slow rotation, $\hat{\tau}_{\mathrm{rad}}\hat{\tau}_{\mathrm{drag}} \sim \mathcal{O}(1)$), the maximum and minimum thermal (height) perturbations $\eta$ lie at the equator and are close to the substellar and antistellar points.
As the timescales get longer (at rapid rotation, $\hat{\tau}_{\mathrm{rad}}\hat{\tau}_{\mathrm{drag}} \sim \mathcal{O}(100)$), the height extrema shift westward and off the equator while cyclonic and anticyclonic gyres develop around them. In the shallow water solutions, the shift between on-equator and off-equator maxima is gradual instead of abrupt because the solutions are linearized and exclude non-linear feedback. Nevertheless, Figure \ref{SWE}a shows the linearized shallow water solutions transition from a large day-night gradient to a more banded structure with an off-equator height maximum, crudely matching the transition from slow over rapid to bat rotators in our GCM simulations (see Figure \ref{batpattern-2}).

What combination of $\hat{\tau}_{\mathrm{rad}}$ and $\hat{\tau}_{\mathrm{drag}}$ is necessary to produce `bat-like' behavior with off-equator hot spots in the shallow water equations? Figure \ref{SWE}a suggests a threshold of $\hat{\tau}_{\mathrm{rad}}\hat{\tau}_{\mathrm{drag}} \sim \mathcal{O}(10)$. To quantify this threshold, we numerically analyzed the analytical solutions from \citet{showman_equatorial_2011} to find the latitude $\widetilde{y}$ at which the height field $\eta$ reaches its maximum. Figure \ref{SWE}c shows $\widetilde{y}$ as a function of $\hat{\tau}_{\mathrm{rad}}$ and $\hat{\tau}_{\mathrm{drag}}$. The plot shows $\widetilde{y}(\hat{\tau}_{\mathrm{rad}},\hat{\tau}_{\mathrm{drag}})$ is approximately symmetric about the diagonal, so it only depends on the product $\hat{\tau}_{\mathrm{rad}} \hat{\tau}_{\mathrm{drag}}$. We take $\widetilde{y}=1$ as the critical value which distinguishes between solutions with on-equator versus off-equator height maxima (black line in Figure \ref{SWE}c).
This boundary can be fitted as $\hat{\tau}_{\mathrm{rad}}\hat{\tau}_{\mathrm{drag}}\approx 13$. Converting the boundary to dimensional quantities, the shallow water equations therefore transition towards bat-like solutions at a threshold orbital period of
\begin{equation}
    \mathrm{P}_\mathrm{bat}\approx\tau_{\mathrm{rad}}\tau_{\mathrm{drag}}\frac{U_\mathrm{wave}}{a}.
    \label{swe-boundary}
\end{equation}
Note, no numerical factors remain in the equation because $\hat{\tau}_{\mathrm{rad}}\hat{\tau}_{\mathrm{drag}}\approx 13$ is cancelled almost exactly by a factor of $4\pi$ which arises when converting from $\beta$ to $P$.

Next, we evaluate the value of $\tau_{\mathrm{rad}}\tau_{\mathrm{drag}}\frac{U_\mathrm{wave}}{a}$ for our GCM simulations. To do so we assume an internal gravity wave with phase speed $U_\mathrm{wave} \sim 20\ \mathrm{m s^{-1}}$, consistent with typical gravity wave speeds in Earth's tropics \citep{kiladis_convectively_2009,vallis_atmospheric_2017}. 
The radiative timescale is equal to $\tau_{\mathrm{rad}}=\frac{c_pp}{4\sigma T_e^3g}$,
where $c_p \approx 1004.64\ \mathrm{J/(kg\cdot K)}$ is the isobaric specific heat of air, $p$ is the surface pressure, $\sigma$ is the Stefan–Boltzmann constant, $T_e$ is the effective radiating temperature, and $g$ is the surface gravity.
To evaluate $\tau_{\mathrm{rad}}$ we use $T_e$ derived from the GCM's global-mean outgoing longwave radiation; we find $\tau_{\mathrm{rad}}$ is about 25-30 days ($2.5 \times 10^6$ s).
Assuming drag in the GCM is mostly from surface friction, $\tau_{\mathrm{drag}}=\frac{\mathcal{U}\rho h}{f}$,
where $\mathcal{U}$ is the horizontal wind at 970 hPa, $\rho$ is the air density at at 970 hPa, $h$ is the thickness of the planetary boundary layer, and $f$ is the zonal surface stress. Evaluating these parameters using GCM output, we find a drag timescale of about $5 \times 10^4$ s, or less than 1 day. For comparison, typical radiative and drag timescales for Earth are 30 days and $5-10$ days \citep{vallis_atmospheric_2017}, and idealized GCM simulations often use a drag timescale of 1 day \citep{held_proposal_1994, penn_atmospheric_2018}.

Plugging these values into Equation \ref{swe-boundary}, the threshold towards bat-like rotation predicted by the linearized shallow water equations is
\begin{equation}
    \mathrm{P}_\mathrm{bat} \approx 4 ~\mathrm{days}.
    \label{gcm-boundary}
\end{equation}
This prediction is larger than the threshold $P \sim 1$ days we numerically found in the GCM by a factor of four; that is, both values agree to within an order of magnitude.
To explain the remaining mismatch between theory and GCM, note that the GCM includes additional mechanisms which are missing from the linearized shallow water equations. In the GCM the transition from rapid rotation to bat rotation is abrupt (see Figure \ref{batdistribution}), and thus is likely related to nonlinear dynamical terms. Similarly, the GCM includes feedbacks from condensation and clouds, and the solutions from \citet{showman_equatorial_2011} restrict the shallow water equations to an equatorial beta-plane, whereas the GCM incorporates full spherical geometry. Any of these factors could create a mismatch by a factor of a few. We therefore consider the shallow water equations to be in qualitative agreement with our far more sophisticated GCM simulations.

As a sanity check, we also analyzed the predicted transition between day-night dominated solutions to solutions with eastward hot spot offsets in the shallow water equations (Case A versus B in Figure \ref{SWE}a). The transition towards eastward hot spot offsets occurs at $\hat{\tau}_{\mathrm{rad}}\hat{\tau}_{\mathrm{drag}} \sim \mathcal{O}(1)$, or one order of magnitude slower rotation that the transition to bat-like behavior. This qualitatively matches the transition between slow and rapid rotators in the GCM, which occurs at an orbital period of $P \sim 20$ days.

What does our result tell us about dynamical regimes on rocky planets versus hot Jupiters? 
We use Equation \ref{swe-boundary} to scale from $P \sim 1$ days for rocky planets to gas giants. Largely by virtue of being hot, the radiative timescale on hot Jupiters is shorter than on Earth-like planets. Using representative hot Jupiter values ($T_{eq} \sim 1500$ K, $p \sim 1$ bar, $g \sim 20$ m s$^{-2}$, c$_p$ corresponding to H$_2$), $\tau_{\mathrm{rad}} \approx 1$ day. Drag timescales for hot Jupiters are widely uncertain, ranging from $\mathcal{O}(10^8)$ s for magnetic drag \citep{perna_magnetic_2010}, down to $\mathcal{O}(10^5)$ s for frictional dissipation associated with shear instabilities \citep{koll_atmospheric_2018}; here we assume a $\tau_{\mathrm{drag}} \approx 1$ day. Finally, we assume a wave speed smaller than the speed of sound by about a factor of two, $U_\mathrm{wave} \sim 10^3$ m s$^{-1}$, and a radius equal to that of Jupiter. Using these values and $P \sim 1$ days for rocky planets, we find hot Jupiters should transition towards bat-like rotation at an orbital period of $P \sim 0.3$ days (equivalently, at an orbital distance less than 0.01 AU). Such high rotation rates and small orbits are generally not accessible to hot Jupiters, as gas giants are rare on ultra-short orbits around main sequence stars \citep{winn_kepler-78_2018}. The bat rotation regime might thus only be accessible to rocky planets.

\section{Inner Edge of Habitable Zone Shaped by Bat Rotation Regime}
\label{inneredge}

Up to now our results show that habitable white dwarf planets on ultrashort orbits exhibit novel atmospheric dynamics. How do these dynamics influence the habitable zone around white dwarfs?

\subsection{Runaway Greenhouse Limit}
\label{sect:RGHL}
To investigate the inner edge of the habitable zone, we adopt a methodology similar to that of \cite{kopparapu_inner_2016}. We move planets closer to the star by simultaneously increasing stellar flux and rotation rate until simulations become unstable. The last converged simulation is taken as a proxy for the Runaway Greenhouse Limit (RGHL; for the full list of converged \& nonconverged simulations see Appendix Figure \ref{appenRGHL}).

\begin{figure}[htbp]
    \centering
    \includegraphics[width=0.6\textwidth]{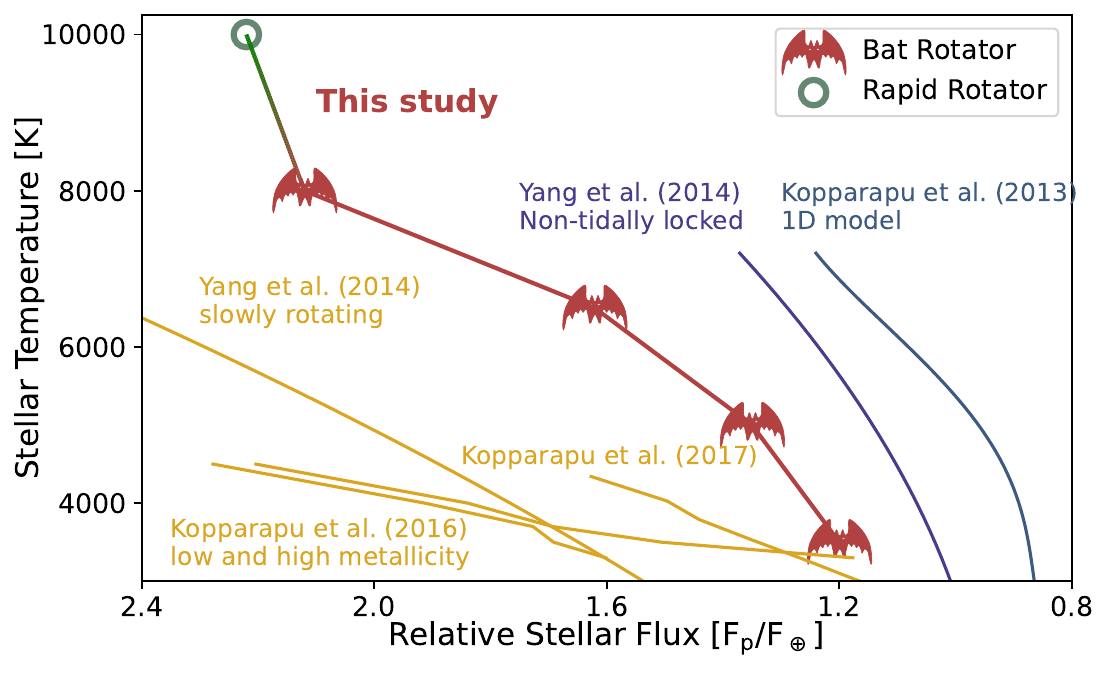}
    \caption{The Runaway Greenhouse Limit (RGHL) around white dwarfs, and comparison to previous studies. 
    X-axis shows the maximum stellar flux at the inner edge of the habitable zone relative to Earth's solar constant, y-axis shows stellar effective temperature.
    The red-green line shows the estimated RGHL for white dwarfs, where bat and circle symbols represent bat rotators and rapid rotators (same as in Figure \ref{batdistribution}). Other colored lines show previous work: RGHL for slowly rotating tidally and locked planets orbiting M dwarfs \citep[yellow;][]{yang_stabilizing_2013,kopparapu_inner_2016, kopparapu_habitable_2017}, RGHL for rapidly rotating and non-tidally locked planets \citep[purple;][]{yang_strong_2014}, and RGHL based on 1D calculations \citep[blue;][]{kopparapu_habitable_2013}.}
    \label{RGHL}
\end{figure}

Figure \ref{RGHL} shows the estimated RGHL as a function of stellar effective temperature. In line with previous habitability studies \citep[e.g.,][]{kopparapu_habitable_2013,yang_strong_2014,kopparapu_habitable_2017,haqq-misra_demarcating_2018}, the RGHL for white dwarfs moves closer to the star at higher stellar temperature. This is because hotter stars give off bluer radiation, which increases the planetary albedo via stronger Rayleigh scattering and weaker atmospheric absorption in the near-infrared \citep[e.g.,][]{kopparapu_inner_2016}. In addition, the RGHL for white dwarfs is intermediate between two limits studied in previous work; it is closer to the star than the RGHL for non-tidally locked planets (red-green versus purple line), but it is further away than the RGHL for tidally locked planets around M dwarfs (red-green versus yellow lines). Note that for most white dwarfs the hottest converged simulation is a bat rotator; only for a 10000 K white dwarf is the hottest converged simulation a rapid rotator (bat versus circle symbols in Figure \ref{RGHL}).

Why does the RGHL for white dwarfs lie in-between that for rapidly rotating non-tidally locked planets and that for slowly rotating tidally locked planets? On the one hand, being tidally locked promotes higher dayside cloud cover, which increases planetary albedo \citep{yang_stabilizing_2013}. On the other hand, once a planet is tidally locked, a shorter orbital period tends to reduce dayside cloud cover, which reduces planetary albedo \citep{yang_strong_2014,kopparapu_inner_2016}. White dwarf planets are both tidally locked and rotate more rapidly than M dwarf planets, matching the sequence in Figure \ref{RGHL}. Our results are thus consistent with previous work on clouds and exoplanet habitability.

Next, we find that the onset of bat rotation helps extend the habitable zone. Figure \ref{RGHL} shows the white dwarf RGHL generally moves inward for hotter stars, but the trend shows a distinct break at 10000 K. This point corresponds to a shift from the hottest converged simulation being a bat rotator to being a rapid rotator (see symbols in Fig.~\ref{RGHL}), suggesting that the onset of bat rotation triggers a stabilizing cloud feedback.

\begin{figure}[htbp]
    \centering
    \includegraphics[width=0.9\textwidth]{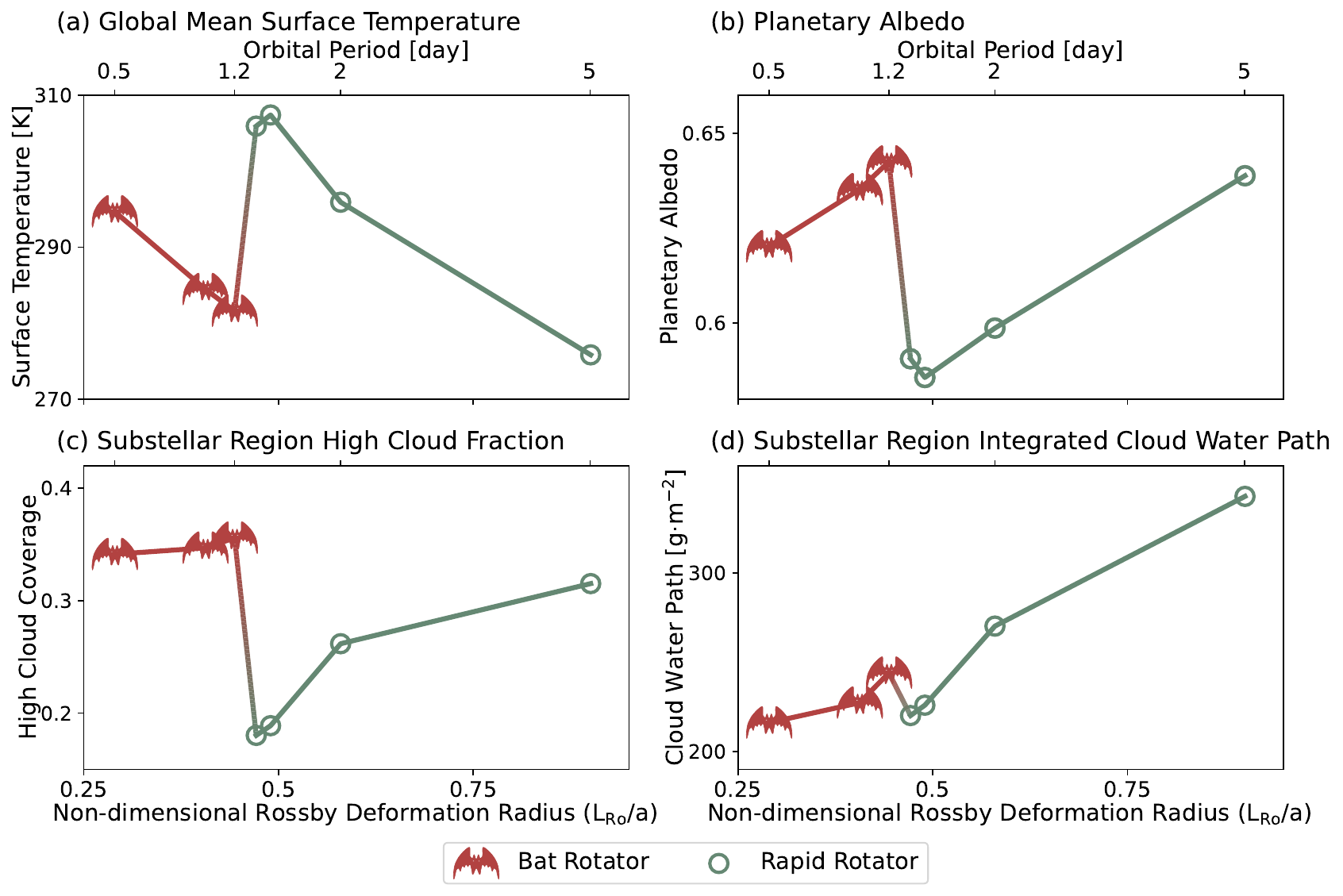}
    \caption{The transition to bat rotation is associated with an abrupt increase in cloud cover, which increases planetary albedo. Simulations shown assume a fixed stellar spectrum and instellation (10000 K, 2.12 times Earth's insolation), and only vary rotation period. (a) Global mean surface temperature. (b) Planetary albedo. (c) High cloud fraction in the substellar region (within 30$^\circ$ of the substellar point) (d) Integrated cloud water path in the substellar region. Bat and circle symbols represent bat rotators and rapid rotators (same as in Figure \ref{batdistribution}).}
    \label{cooling}
\end{figure}

To investigate this cloud feedback in detail, we performed another set of idealized simulations for a 10000 K host star in which we vary rotation rate while keeping instellation fixed. Varying rotation from 5 days down to 0.5 days, we find the same trend as in prior studies; more rapid rotation tends to reduce dayside cloud cover and planetary albedo \citep{yang_strong_2014,kopparapu_inner_2016}. However, Figure \ref{cooling} shows that the general trend is interrupted by an abrupt jump at about 1.2 days, when the atmospheric circulation regime changes from rapid rotation to bat rotation. Global mean surface temperature abruptly cools by $\sim$30 K, while planetary albedo increases by $\sim$10\% (Fig.~\ref{cooling}a,b). The increase in albedo is linked to a jump in high cloud fraction and cloud water path inside the substellar region. Both increase at 1.2 days, before decreasing again at even higher rotation (Fig.~\ref{cooling}c,d). Therefore, white dwarf planets exhibit the same general cloud feedback as M dwarf planets, in which higher rotation reduces cloud reflection and tends to shrink the habitable zone, but the shift from rapid rotation to bat rotation (not accessible to most habitable M dwarf planets) triggers a one-time increase in cloud reflection, expanding the habitable zone.

To estimate how clouds affect the overall width of the white dwarf habitable zone, we combined our GCM simulations for the inner edge of the habitable zone with prior 1D model results for the outer edge of the habitable zone \citep[see Equation \ref{outer} below, from][]{kopparapu_habitable_2013}.
Over the range of stellar temperatures for which Equation \ref{outer} is valid, we find that our GCM calculations increase the width of the habitable zone in instellation space by 50-60\% compared to estimates based on 1D models \citep{agol_transit_2011,kopparapu_habitable_2013}.

\subsection{Moist Greenhouse Limit}
Next, we investigate whether the habitable zone around white dwarfs is sensitive to gradual water loss via partial atmospheric escape to space, the so-called Moist Greenhouse. Habitable planets constantly lose water because H$_2$O in the upper atmosphere photo-dissociates into H and O. The lighter H preferentially escapes to space, gradually drying out the planet. A planet exceeds the Moist Greenhouse Limit (MGHL) if it loses an Earth ocean's worth of water within 4.5 Gyr; this occurs when the volume mixing ratio of stratospheric water vapor is higher than about $3\times 10^{-3}$ \citep{kasting_habitable_1993}. In practice, we evaluate this criterion using the global-mean specific humidity at the top of model. One can also define the MGHL in terms of the top-of-model specific humidity in the substellar region only \citep{kopparapu_inner_2016}; both MGHL definitions are consistent to within one order of magnitude so here we use the global-mean definition.

\begin{figure}[htbp]
    \centering
    \includegraphics[width=0.65\textwidth]{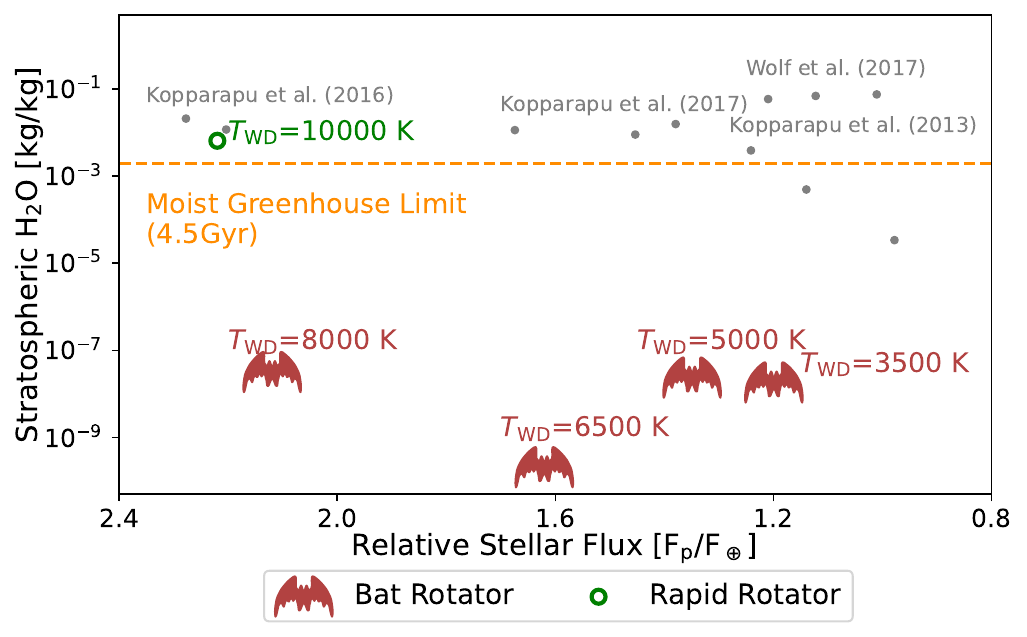}
    \caption{Bat rotators have exceptionally dry upper atmospheres, which suppresses the Moist Greenhouse. X-axis shows stellar flux of the last converged simulation, y-axis shows the top-of-model specific humidity. The orange dashed line indicates a specific humidity of $1.93 \times 10^{-3}$ kg/kg, corresponding to a volume mixing ratio of $3 \times 10^{-3}$. Grey dots show results from previous studies for planets orbiting main-sequence stars \citep{kopparapu_habitable_2013,kopparapu_inner_2016,wolf_constraints_2017,kopparapu_habitable_2017}; bat and circle symbols represent bat rotators and rapid rotators, same as in Figure \ref{batdistribution}; and $T_\mathrm{WD}$ is the effective temperature of the white dwarf.}
    \label{MGHL}
\end{figure}

The top-of-model specific humidities of the last converged (i.e., hottest) simulations are shown in Figure \ref{MGHL}. Unexpectedly, white dwarf planets with a host star cooler than 10000 K never exceed the MGHL. This is because bat rotators have extremely dry upper atmospheres. Whereas hot main sequence planets considered in previous work featured stratospheric specific humidities ranging from 10$^{-1}$ to 10$^{-4}$ kg/kg \citep{kopparapu_habitable_2013,kopparapu_inner_2016,wolf_constraints_2017,kopparapu_habitable_2017}, and thus typically exceeded the Moist Greenhouse Limit, the hottest bat rotators have top-of-model specific humidities ranging from a dry 10$^{-7}$ kg/kg (drier than Earth's stratosphere) down to an exceptionally dry 10$^{-10}$ kg/kg.

\begin{figure}
    \centering
    \includegraphics[width=0.8\textwidth]{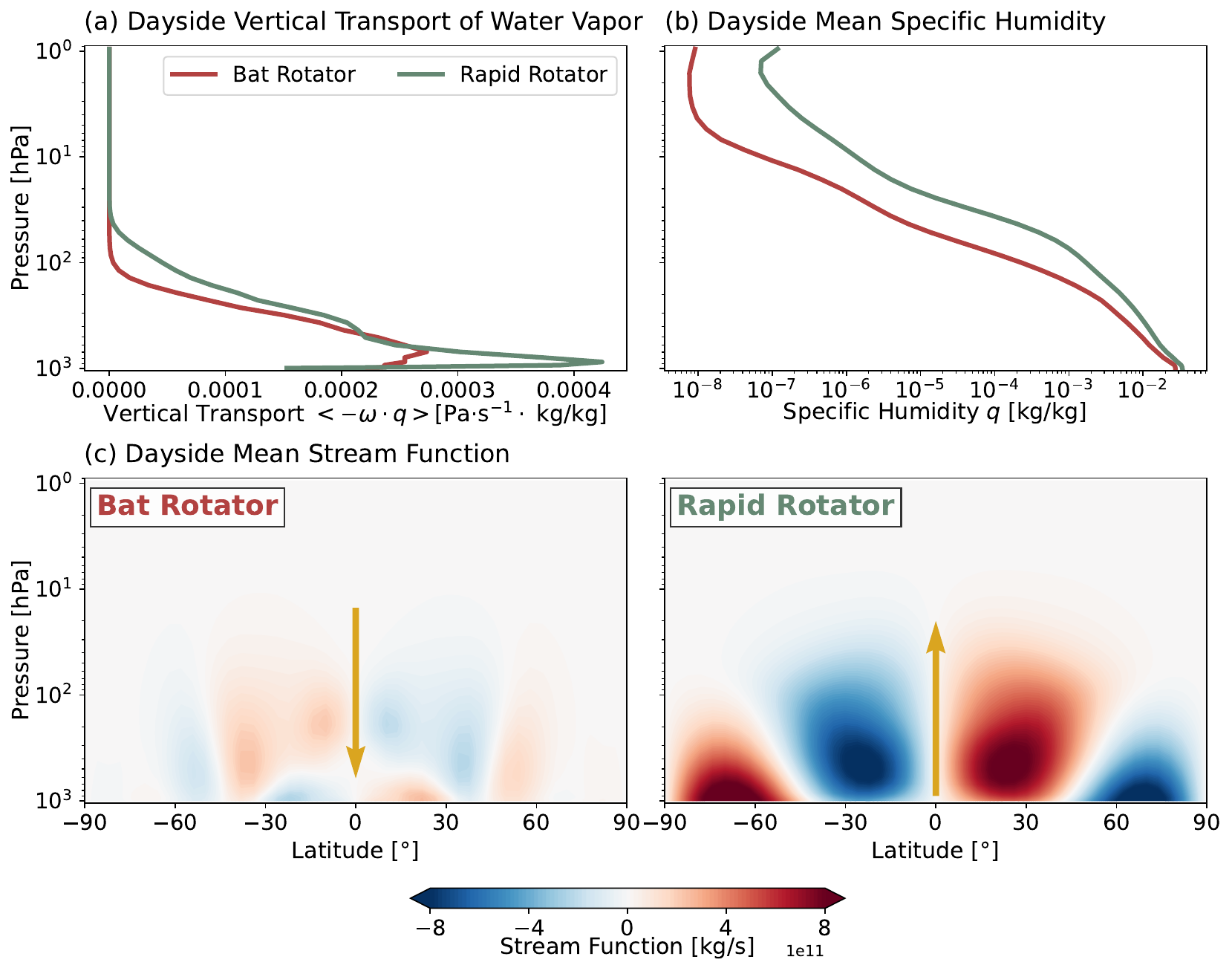}
    \caption{Bat rotators have drier upper atmospheres and less efficient vertical water transport than rapid rotators. (a) Dayside mean vertical transport of water vapor; a larger $<-\omega\cdot q>$ represents stronger upward transport of water vapor. (b) Vertical profile of dayside mean specific humidity. (c) Zonal mean stream function on the dayside hemisphere; red shading (positive) indicates clockwise circulation, blue shading (negative) indicates counter-clockwise circulation. Yellow arrows indicate the direction of vertical transport by the mean circulation at the equator. The bat rotator and the rapid rotator are orbiting a 10000 K white dwarf and receiving 2.12 times Earth's insolation. The orbital period of the bat rotator and the rapid rotator are 0.5 day and 2 days, identical to those in Figure \ref{batandrapid}.}
    \label{dry}
\end{figure}

The dry upper atmospheres of bat rotators are associated with a reversal in the dayside atmospheric circulation and the formation of anti-Hadley cells, which suppress vertical transport of water vapor. Figure \ref{dry} compares dayside specific humidity, vertical water vapor transport, and meridional streamfunction for a bat rotator versus a rapid rotator; the shown simulations were chosen because they have almost identical global mean surface temperatures and surface specific humidities. Despite the similar lower boundary, above 500 hPa the dayside mean specific humidity on the bat rotator is approximately 10 times less than on the rapid rotator, while vertical water vapor transport is a factor of 2-10 less (Figure \ref{dry}a,b).

To explain the difference in vertical water vapor transport, Figure \ref{dry}c displays the dayside zonal mean stream functions. On the rapid rotator the dayside circulation consists of air rising at the equator and sinking in the mid-latitudes, akin to a Hadley cell. Surprisingly, on the bat rotator the circulation is flipped; above 600 hPa, air sinks at the equator and rises in the mid-latitudes, forming an anti-Hadley cell \citep{charnay_3d_2015}. The anti-Hadley cell is consistent with the unique surface temperature structure of the bat rotator. Its hottest temperatures are located in the mid-latitudes, which reverses the meridional temperature gradient at low latitudes (Figures \ref{batpattern-1},\ref{batandrapid}). Any thermally direct meridional (Hadley-like) circulation thus also has to reverse its direction. This helps to suppress deep convection at the equator and explains the ineffective vertical transport of moist air to the upper atmosphere. In addition, the stream function on the bat rotator is weaker than that of the rapid rotator, implying generally weaker vertical winds.

Overall, white dwarf planets are thus unlikely to experience the Moist Greenhouse; in contrast to most main sequence planets, these planets directly jump from a warm and temperate state into the Runaway Greenhouse (see Figure \ref{MGHL}). Note, \cite{kopparapu_inner_2016} also found that planets orbiting very low mass M dwarfs can skip the Moist Greenhouse, because these stars emit redder radiation which causes strong shortwave absorption and suppresses convection near the substellar point. In contrast, our results here hold over a wide range of stellar effective temperatures, highlighting the importance of planetary rotation for white dwarf planets.

\clearpage

\section{Discussion}
\label{discussion}
\subsection{Significance of the Bat Rotation Regime}

\begin{figure}[t]
    \centering
    \includegraphics[width=0.65\textwidth]{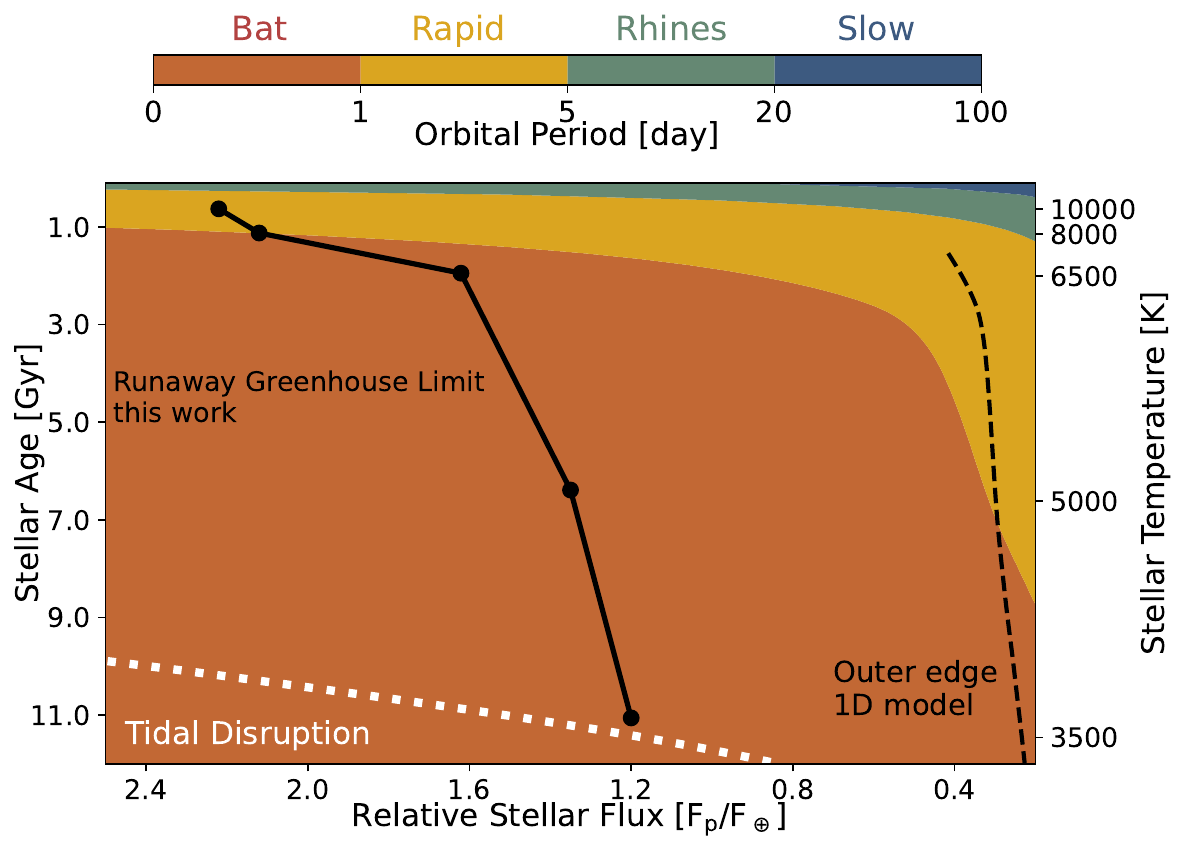}
    \caption{Estimated rotation regimes inside the habitable zone of white dwarfs with different stellar temperatures, as a function of relative stellar flux. Colors denote different rotation regimes. Black solid line shows the inner edge of the habitable zone (RGHL, Section \ref{inneredge}); black dashed line shows the outer edge of the habitable zone \citep[Maximum Greenhouse Limit from][]{kopparapu_habitable_2013}; and white dotted line shows the tidal disruption limit. The stellar mass of the white dwarf is fixed at 0.6 $M_\Sun$.}
    \label{regimes}
\end{figure}

The bat rotation regime is important near the inner edge of the habitable zone around white dwarfs, but what about the rest of the habitable zone? In this section we argue most habitable planets around white dwarfs should also be bat rotators.

To assign different dynamical regimes to planets as a function of orbital period $P$ we use criteria based on \citet{haqq-misra_demarcating_2018}: planets with $P > 20$ days are assumed to be slow rotators, planets with $5\text{~days}< P < 20\text{~days}$ are Rhines rotators, planets with $1\text{~days}< P < 5\text{~days}$ are rapid rotators, and planets with $P < 1$ day are bat rotators.
To translate stellar effective temperature into stellar age, we use a model\footnote{https://github.com/SihaoCheng/WD\_models} for white dwarfs with thin hydrogen envelopes adopted from \cite{bedard_spectral_2020}. To estimate the outer edge of the habitable zone, we use a fit for the Maximum Greenhouse Limit from 1D model calculations \citep{kopparapu_habitable_2013},
\begin{equation}
    S_{\text{eff}} = 0.3438 + 5.8942\times10^{-5}T_\star + 1.6558\times10^{-9}T_\star^2 - 3.0045\times10^{-12}T_\star^3 - 5.2983\times10^{-16}T_\star^4,
    \label{outer}
\end{equation}
\begin{equation}
    T_\star=T_{\text{WD}}-5780, \qquad 2600 K \leq T_{\text{WD}} \leq 7200 K
\end{equation}
where $S_{\text{eff}}$ and $T_{\text{WD}}$ are the instellation at the outer edge of the habitable zone and the effective temperature of the white dwarf.

As white dwarfs age and cool, the habitable zone moves closer to the star; habitable planets around very old white dwarfs are thus in danger of tidal disruption. To estimate the tidal disruption boundary, we use an approximate solution of the Roche limit from \cite{lissauer_fundamental_2019} valid for tidally locked liquid planets,
\begin{equation}
    d_\text{tidal} \approx 2.456 a_{\text{WD}} \left( \frac{\rho_{\text{WD}}}{\rho_p}\right)^\frac{1}{3}.
\end{equation}
Here $d_\text{tidal}$ is tidal disruption distance, $a_{\text{WD}}$ and $\rho_{\text{WD}}$ are the white dwarf radius and mean density, and $\rho_p$ is the planet's mean density. Assuming a white dwarf with $M_\text{WD}=0.6 M_\Sun$ and $\rho_p$ equal to Earth's density, the tidal disruption distance for white dwarfs is $d_\text{tidal}=0.0037$ AU. The corresponding instellation $S_\text{tidal}$ at this distance is, 
\begin{equation}
    S_\text{tidal} = S_\oplus\left( \frac{T_{\text{WD}}}{T_\odot} \right)^4 \left(\frac{a_{\text{WD}}}{a_\odot}\right)^2 \left(\frac{1 \text{AU}}{d_\text{tidal}}\right)^2,
\end{equation}
where $S_\oplus$ is Earth's solar constant, and $a_{\text{WD}}$ and
$a_\odot$ are the radius of the white dwarf and Sun.

Figure \ref{regimes} shows that most of the habitable zone around white dwarfs corresponds to the bat rotation regime. The only overlap with the rapid rotation regime occurs for white dwarfs younger than 3 Gyr.
In addition, tidal disruption cuts off the habitable zone around stars older than about 11-12 Gyr. This upper limit for the oldest white dwarf systems that can still be habitable is about 50\% more than the value estimated based on 1D models by \citet{agol_transit_2011}, consistent with the fact that our white dwarf habitable zone based on 3D GCMs is wider than that based on 1D models (see Figure \ref{RGHL}).
Note that one could in principle sustain habitable conditions around white dwarfs even older than 12 Gyr, but doing so would require unconventional conditions that go beyond the conventional habitable zone \citep[e.g., H$_2$-rich atmospheres; ][]{pierrehumbert_hydrogen_2011}.

\subsection{Observation Implications}

\begin{figure}[htbp]
    \centering
    \includegraphics[width=0.95\textwidth]{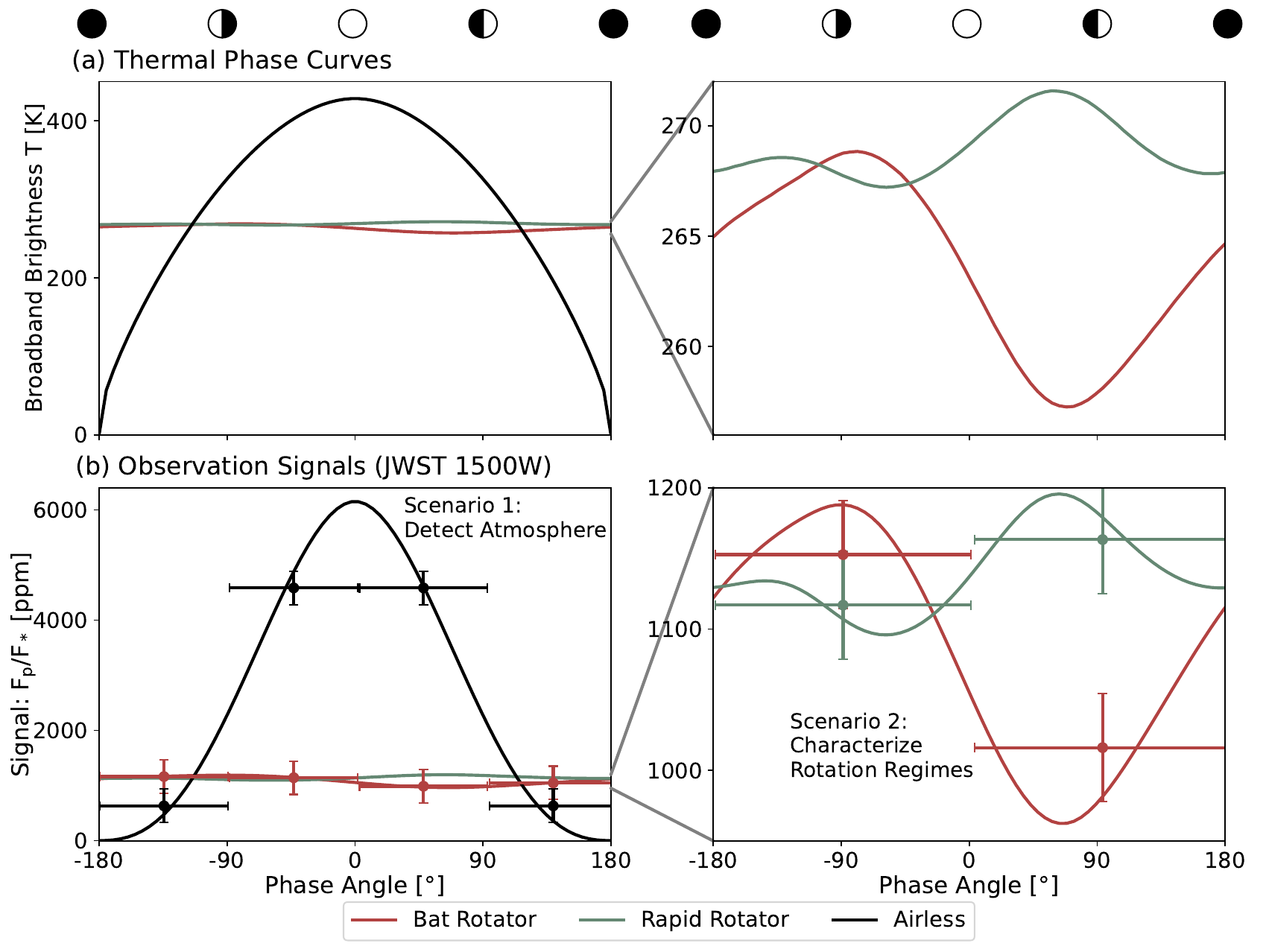}
    \caption{(a) Thermal phase curves of a a bat rotator, a rapid rotator, and an airless planet, shown as $15 \mu$m brightness temperatures. (b) Observable signals compared to a photon noise estimate; error bars show 1$\sigma$ photometric precision with JWST MIRI F1500W. Scenario 1 (Detect Atmosphere) assumes a white dwarf at 25 pc and 3 hours per exposure. Scenario 2 (Characterize Rotation Regimes) assumes a white dwarf at 10 pc and 6 hours per exposure. The bat rotator and the rapid rotator are configured the same as in Figure \ref{batandrapid}. Phase curves of the bat rotator and the rapid rotator are based on GCM simulations and Equation \ref{phasecal}, for the airless planet we assume zero albedo.}
    \label{phasecurve}
\end{figure}

Can our theoretical predictions be tested? Previous work proposed using JWST transmission spectra to search for atmospheric absorption features and biosignatures on habitable white dwarf planets  \citep{agol_transit_2011,kaltenegger_white_2020}. Here we consider a complementary proposal, namely using JWST thermal observations to identify whether white dwarf planets have atmospheres in the first place \citep{selsis_thermal_2011,koll_identifying_2019,mansfield_identifying_2019,malik_analyzing_2019}, and, with a larger investment of observing time, to empirically characterize the atmospheric dynamics regime of these planets.

Our calculation follows previous thermal phase curve calculations \citep{cowan_inverting_2008,cowan_thermal_2012,koll_deciphering_2015,koll_scaling_2022}: we assume a white dwarf-planet system in a transiting geometry, with the planet's orbit viewed edge-on. The planet emits thermal radiation isotropically. The intensity $I$ at the top-of-atmosphere is 
\begin{equation}
    I_{\mathrm{TOA}}(\phi, \theta)=\frac{F_{\mathrm{LW,TOA}}^\uparrow(\phi, \theta)}{\pi},
\end{equation}
where $F_{\mathrm{LW,TOA}}^\uparrow$ is the GCM's flux of outgoing longwave radiation, while $\phi$ and $\theta$ are longitude and latitude on the planet. The planet's thermal flux received by a distant observer as a function of orbital phase $\xi$ is (see Appendix \ref{thermalflux})
\begin{equation}
    F(\xi)=\frac{\pi a^2}{d^2}\frac{\int_{-\frac{\pi}{2}}^{\frac{\pi}{2}}d\theta\int_{-\xi-\frac{\pi}{2}}^{-\xi+\frac{\pi}{2}}I_{\mathrm{TOA}}(\phi,\theta)\cos(\phi+\xi)\cos^{2}\theta d\phi}{\int_{-\frac{\pi}{2}}^{\frac{\pi}{2}}d\theta\int_{-\xi-\frac{\pi}{2}}^{-\xi+\frac{\pi}{2}}\cos(\phi+\xi)\cos^{2}\theta d\phi},
    \label{phasecal}
\end{equation}
where $a$ is planet radius, and $d$ is the distance between observer and planetary system. We express thermal phase curves either in terms of the planet-star flux ratio $F_p/F_*$, or using the planet's effective brightness temperature.

We define the observable signal for a planet's phase curve as as the difference between maximum and minimum flux (although shift in phase curve is another observable signal, see below), computed as
\begin{equation}
    \mathrm{Signal}=\left(\frac{a_\mathrm{p}}{a_*}\right)^2
\frac{\int_{\lambda_1}^{\lambda_2}
\left[\max(I_{\mathrm{p},\lambda}(\xi))-\min(I_{\mathrm{p},\lambda}(\xi))\right]d\lambda}
{\int_{\lambda_1}^{\lambda_2}B_{*,\lambda}d\lambda},
    \label{eqn:signal}
\end{equation}
where $a_\mathrm{p}$ and $a_*$ is the radius of planet and star,
($\lambda_1$,$\lambda_2$) is the telescope bandpass over which we average, $F_*$ is the flux from the host star (assumed to emit as a blackbody), and $B_*$ is the intensity or brightness of the star.

To model the observational noise, we assume photon noise. Recent JWST observations with MIRI-LRS indicate a noise floor of at most $\sim25$ ppm \citep{bouwman_spectroscopic_2023}; this is less than the $\mathcal{O}(1000)$ ppm signal for atmospheric detection and $\mathcal{O}(100)$ ppm signal for atmospheric characterization of white dwarf planets. The flux from the planetary system is dominated by the star \citep{cowan_characterizing_2015}, so the observed number of photons equals
\begin{equation}
    N_{\mathrm{photon}}=\frac{\pi\tau\Delta t}{hc} \left( \frac{a_*D}{2d}\right)^2 \int_{\lambda_1}^{\lambda_2}B_*(\lambda)\lambda d\lambda,
\end{equation}
where $\tau$ is telescope throughput, $\Delta t$ is the integration time, $D$ is telescope diameter, and $B_*$ is the star's flux which we assume follows Planck's law. In the Poisson limit of large $N_{\mathrm{photon}}$, the 1$\sigma$ precision is approximately $\sqrt\frac{2}{N_{\mathrm{photon}}}$, where the factor of $\sqrt{2}$ originates from the fact that the phase curve amplitude is measured as the difference between two flux measurements (Equation \ref{eqn:signal}). The signal-to-noise ratio for thermal phase curves of temperate planets is highest in the mid-infrared, so we assume observations are performed with JWST-MIRI F1500W (i.e., at 15$\mu$m). For the white dwarf we assume an effective temperature of 10000 K and the same stellar radius as in the GCM calculations ($a = 0.012 a_\odot $).

We consider three planetary scenarios: an atmosphere-less blackbody planet, a rapid rotator, and a bat rotator. The latter two are both habitable and host a 1 bar N$_2$-H$_2$O atmosphere, with their climates and phase curves simulated by the GCM (to simplify the comparison we use the two idealized simulations with identical stellar flux from Figure \ref{batandrapid}).
Figure \ref{phasecurve}a shows the three planet scenarios result in very different phase curves. The airless planet has a large phase curve amplitude and is symmetric about the secondary eclipse (phase angle 0$^\circ$). In comparison, the two atmospheric scenarios have phase curves that are almost flat compared to an airless planet, due to the atmosphere's effective day-night heat redistribution.

Zooming in on the difference between bat and rapid rotator, we find that they exhibit opposite phase shifts (Figure \ref{phasecurve}a). 
The rapid rotator's phase curve maximum occurs after secondary eclipse, consistent with the fact that its equatorial cloud-free region is west of the substellar point \citep[see Figure \ref{batandrapid} and ][]{haqq-misra_demarcating_2018}.
In contrast, the bat rotator's phase curve maximum occurs before secondary eclipse, consistent with the fact that its equatorial cloud-free region is east of the substellar point (Figure \ref{batandrapid}). Bat rotators can thus be distinguished from rapid rotators based on their hot spot offsets.

How long would it take to measure these differences with JWST? Figure \ref{phasecurve}b shows two observing scenarios. Scenario 1 ``Detect Atmosphere" assumes the system is 25 pc away, the same distance as WD 1856+534b which is the closest known transiting (giant) planet around a white dwarf \citep{vanderburg_giant_2020}. For a bat rotator on a 12h orbit, a single JWST phase curve (12 h total observation time, binned into 4 data points) can measure the phase curve amplitude of an airless planet with >10$\sigma$ confidence. That is sufficient to confidently distinguish between no atmosphere and a thick atmosphere.
Scenario 2 ``Characterize Rotation Regimes" assumes a more optimistic target at 10 pc.
Although 10 pc is closer than any currently known transiting planet around a white dwarf, JWST might also be able to measure phase curves of non-transiting planets around white dwarfs \citep{limbach_new_2022}, which should greatly increase the number of nearby accessible targets. In this case 5 stacked phase curves of the bat rotator (60 h total observation time, binned into 2 data points) would be sufficient to measure the phase curve amplitude of the bat rotator with 3$\sigma$ confidence. That is sufficient to characterize the bat rotator's hot spot offset, and distinguish between a bat rotator and rapid rotator. This observational effort is significantly less than that required to characterize habitable M dwarf planets such as the TRAPPIST-1 planets \citep{lustig-yaeger_detectability_2019}. It arises because white dwarfs are much smaller than M dwarfs, significantly boosting the planet-to-star flux ratio and thus the observable signal \citep[see Eqn.~15 and ][]{kaltenegger_white_2020}. Overall, habitable white dwarf planets are thus attractive targets for characterization via thermal phase curves.

\section{Conclusions}

\label{conclusion}

To summarize, we utilize a 3D GCM to investigate the inner edge of the habitable zone around white dwarfs. Since white dwarfs are compact and have low luminosity, habitable planets around them are likely tidally locked and rotate rapidly, with orbital periods ranging from a few hours to several days. We investigate the atmospheric dynamics of habitable tidally locked planets at such high rotation rates, and their implication for the habitable zone around white dwarfs. Our key results are:
\begin{itemize}
    \item Our GCM simulations show habitable white dwarf planets with ultrashort orbital periods ($P\lesssim 1$ day) enter a new atmospheric dynamics regime that is distinct from atmospheric dynamics regimes on more slowly rotating tidally locked planets around main sequence stars \citep{haqq-misra_demarcating_2018}. We name this regime the bat rotation regime due to its characteristic surface temperature pattern. On bat rotators the hottest surface temperatures move off the equator into the mid-latitudes, while mean equatorial winds reverse sign from superrotation to subrotation.
    \item The transition from rapid rotators to bat rotators at $P\sim1$ day suggestively matches the transition from on-equator to off-equator hot spots in the linearized shallow water equations at $P\sim\mathcal{O}(1)$ day. Future work is needed to understand the transition in detail, which in the GCM is abrupt and likely involves non-linear feedbacks. Our results also suggest the bat rotation state might not be accessible to hot Jupiters. 
    \item In the GCM, the transition to bat rotation increases dayside cloud cover which pushes the Runaway Greenhouse Limit closer to the star. In addition, bat rotators have dayside anti-Hadley cells and weak vertical water transport, which leads to extremely dry upper atmospheres, allowing most white dwarf planets to avoid the Moist Greenhouse. As a result, the white dwarf habitable zone based on our GCM calculations is 50-60\% wider in instellation space than the habitable zone estimated based on 1D models.
    \item For white dwarfs older than 3 Gyr, the bat rotation regime overlaps with most of the habitable zone. Many or most habitable white dwarf planets should thus be bat rotators.
    \item Habitable white dwarf planets are attractive targets for characterization via JWST thermal phase curves. For a white dwarf at 25 pc, 12 hours (one full phase curve) should be sufficient to distinguish a thick atmosphere from an airless planet. For a white dwarf at 10 pc, 60 hours (multiple stacked phase curves) should be sufficient to distinguish between a bat rotator and a rapid rotator.
\end{itemize}

We thank Eric T Wolf for developing ExoCAM and making it publicly available. We are grateful to Jun Yang, Tad Komacek, Xianyu Tan, Lixiang Gu, Xinyi Song, and Xintong Lyu for helpful comments. We thank Aomawa Shields and Pier-Emmanuel Tremblay for sharing white dwarf SEDs. D.D.B.K. acknowledges support from the National Natural Science Foundation of China (NSFC) under grant 42250410318. The simulation outputs used to make figures and tables in this study are available on Zenodo \citep{zhan_novel_2024}.

\appendix 
\section{Appendix: Sensitivity Test}
\subsection{White Dwarf Spectra Energy Distribution}
\label{appen:sensSED}
As discussed in Section \ref{method}, our white dwarf SEDs are either from ExoCAM or based on the 3D NLTE models presented in \cite{tremblay_spectroscopic_2013,tremblay_3d_2015}\footnote{https://warwick.ac.uk/fac/sci/physics/research/astro/people/tremblay/modelgrids}. We find that the exact choice of white dwarf SED has a small but non-zero influence on GCM output.

\begin{figure}[htbp]
    \centering
    \subfloat[]{\includegraphics[width=0.75\textwidth]{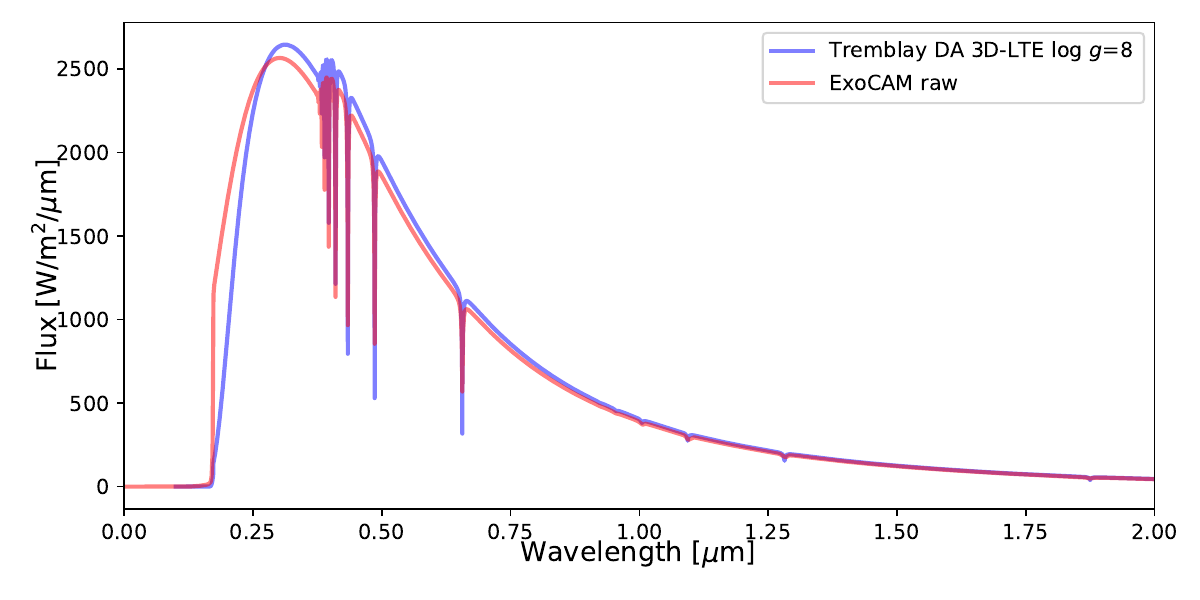}}
    \hfill
    \subfloat[]{\includegraphics[width=0.9\textwidth]{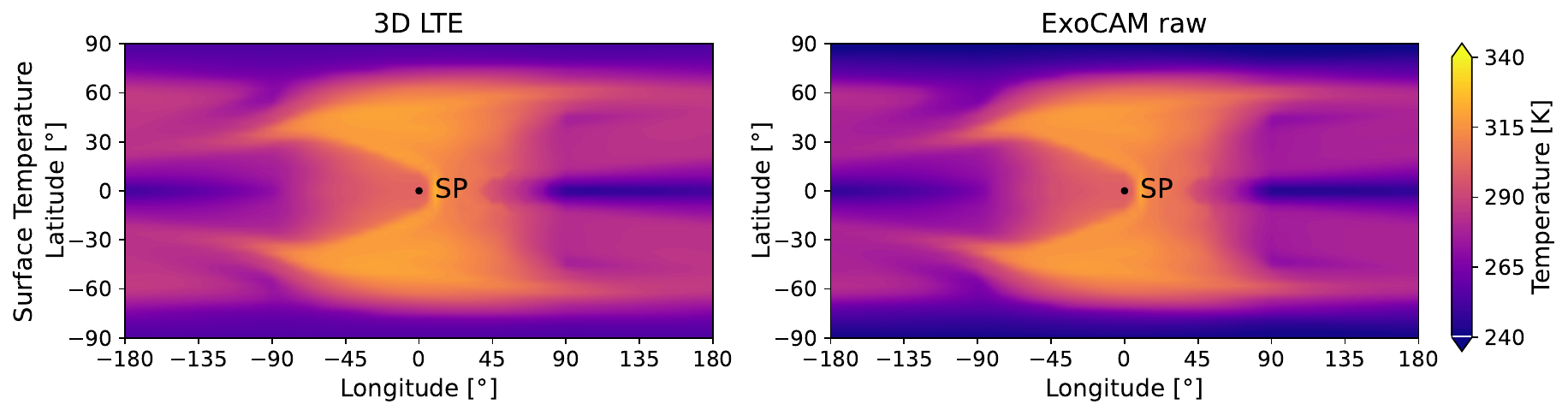}}
    \caption{(a) Two similar spectral energy distributions (SEDs) for an 8000 K white dwarf. Red line shows the default SED from ExoCAM, blue line shows an SED based on a 3D NLTE model (DA spectral type, stellar surface gravity about $10^{10}$ m s$^{-2}$). (b) Surface temperature maps, simulated with ExoCAM. Both simulations are identical except for the stellar SED.}
    \label{sensSED}
\end{figure}

Figure \ref{sensSED} shows a sensitivity test for an 8000 K white dwarf. The top row shows the different SEDs, bottom row shows the resulting GCM surface temperature maps.
We find a moderate difference in global mean surface temperature: 287.0 K with the 3D NLTE model SED versus 282.4 K with the default ExoCAM SED. In addition, Figure \ref{sensSED} shows that the pattern of surface temperature is highly similar between both cases. To be more consistent with other studies that use ExoCAM, we therefore choose to use ExoCAM SEDs whenever possible, and 3D NLTE model SEDs when there are no ExoCAM SEDs provided (i.e., for 3500 K and 6500 K white dwarfs).

\subsection{Horizontal Resolution}
\label{appen:sensHOR}

The Rossby deformation radius, a fundamental length scale for atmospheric dynamics, is smaller on planets with higher rotation rates. We therefore investigate whether the horizontal resolution in ExoCAM is adequate for rapidly-rotating planets.

We consider two white dwarf planets with orbital periods of about 10 hours and 4 hours. Each case is simulated using two different horizontal resolutions, $4^\circ \times 5^\circ$ (corresponding to $\Delta x$ about 450 km$\times$550 km at the equator) and $1.9^\circ \times 2.5^\circ$ ($\Delta x$ about 200 km$\times$250 km at the equator). For the more slowly rotating planet, the $4^\circ \times 5^\circ$ resolution is significantly smaller than the equatorial Rossby deformation radius, $L_{\mathrm{Ro}} \sim 5 \Delta x$, while for the more rapidly rotating planet the $4^\circ \times 5^\circ$ resolution is comparable to the Rossby deformation radius, $L_{\mathrm{Ro}} \sim 2 \Delta x$.

\begin{figure}[htbp]
    \centering
    \subfloat{\includegraphics[width=0.48\textwidth]{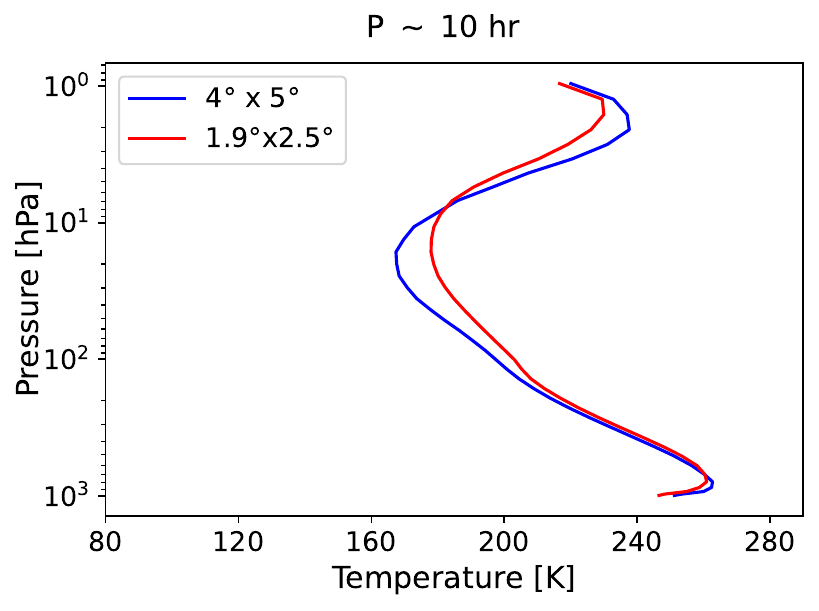}}
    \hfill
    \subfloat{\includegraphics[width=0.48\textwidth]{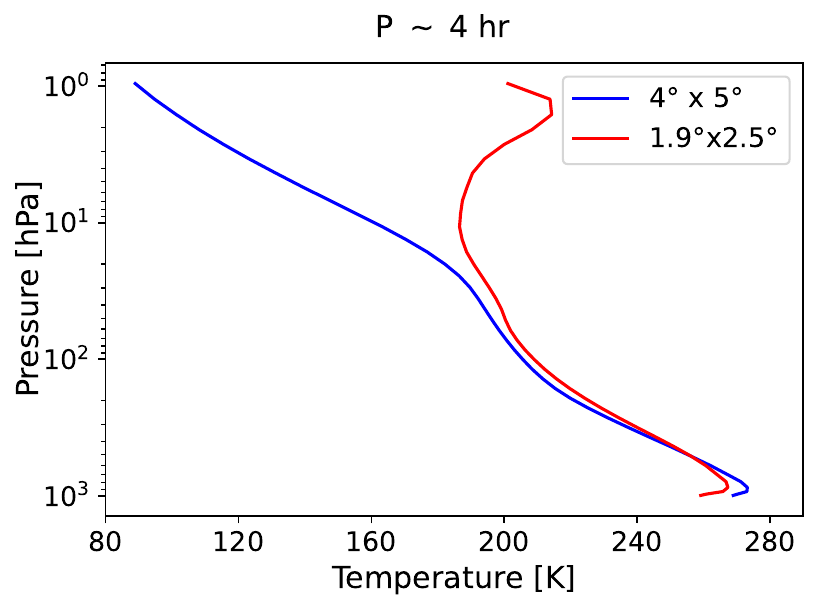}}
    \caption{Sensitivity test on the impact of ExoCAM horizontal resolution. Both panels show global-mean vertical temperature profiles of white dwarf planets receiving the same insolation as Earth. Left panel: 5000 K white dwarf. Right panel: 3500 K white dwarf. Blue lines show lower resolution ($4^\circ \times 5^\circ$), red lines show higher resolution ($1.9^\circ \times 2.5^\circ$).}
    \label{senshorres}
\end{figure}

Figure \ref{senshorres} shows global-mean vertical temperature profiles of the simulations with different horizontal resolutions. We find that, with an orbital period of about 10 hours, increasing the horizontal resolution only has a minor impact on atmospheric temperature structure; temperatures in the upper atmosphere change by about 10 K while temperatures near the surface are essentially the same. However, with a shorter orbital period of about 4 hours, increasing the resolution has a much bigger impact; temperatures in the upper atmosphere change by more than 100 K while temperatures near the surface change by around 5 K.
Our results are qualitatively consistent with the ExoCAM results from \cite{wei_small_2020}, who found that simulations with higher resolution result in drier, warmer upper atmospheres and cooler surfaces. However, \cite{wei_small_2020} only reported a small sensitivity to changes in horizontal resolution, whereas in some of our simulations the sensitivity is large (see Figure \ref{senshorres}, right panel). This is presumably because we are considering much more rapidly rotating planets; the shortest orbital period in \citet{wei_small_2020} was 4.25 days, whereas here we consider orbital periods down to a few hours.

In our GCM simulations we therefore use a resolution of either $4^\circ \times 5^\circ$ or $1.9^\circ \times 2.5^\circ$, as shown in Table \ref{table:configuration}. We do not use resolutions higher than $1.9^\circ \times 2.5^\circ$ because Figure \ref{senshorres} (left panel) suggests that, once the Rossby deformation radius is much larger than the grid scale, any further increase in horizontal resolution does not lead to significant changes in the simulated climate.

\section{Appendix: Simulation Details}
\subsection{Tables}
\label{appen:sensDET}

\begin{table}
\centering
\setlength\tabcolsep{12pt}
\begin{tabular}{ccccc}
\hline\hline
$T_{\mathrm{WD}}$ {[}K{]} & $F_{\mathrm{p}}$/$F_{\oplus}$ & Orbital Period [day] & Horizontal Resolution        & Rotation Regime \\\hline 
\multirow{3}{*}{3500}     & 1.0                           & 0.1474        & $1.9^\circ \times 2.5^\circ$ & bat             \\
                          & 1.2                           & 0.1286        & $1.9^\circ \times 2.5^\circ$ & -               \\
                          & 1.3                           & 0.1211        & $1.9^\circ \times 2.5^\circ$ & (crash)         \\\hline
\multirow{5}{*}{5000}     & 1.0                           & 0.4299        & $4^\circ \times 5^\circ$     & bat             \\
                          & 1.2                           & 0.3749        & $4^\circ \times 5^\circ$     & bat             \\
                          & 1.3                           & 0.3531        & $4^\circ \times 5^\circ$     & bat             \\
                          & 1.35                          & 0.3432        & $4^\circ \times 5^\circ$     & bat             \\
                          & 1.4                           & 0.3340        & $4^\circ \times 5^\circ$     & (crash)         \\\hline
\multirow{5}{*}{6500}     & 1.0                           & 0.9444        & $4^\circ \times 5^\circ$     & -               \\
                          & 1.417                         & 0.7272        & $4^\circ \times 5^\circ$     & bat             \\
                          & 1.5                           & 0.6968        & $4^\circ \times 5^\circ$     & bat             \\
                          & 1.621                         & 0.6574        & $4^\circ \times 5^\circ$     & bat             \\
                          & 1.7                           & 0.6343        & $4^\circ \times 5^\circ$     & (crash)         \\\hline
\multirow{9}{*}{8000}     & 1.0                           & 1.7606        & $4^\circ \times 5^\circ$     & rapid           \\
                          & 1.417                         & 1.3557        & $4^\circ \times 5^\circ$     & rapid           \\
                          & 1.519                         & 1.2868        & $4^\circ \times 5^\circ$     & rapid           \\
                          & 1.621                         & 1.2256        & $4^\circ \times 5^\circ$     & rapid           \\
                          & 1.723                         & 1.1708        & $4^\circ \times 5^\circ$     & rapid           \\
                          & 1.965                         & 1.0609        & $4^\circ \times 5^\circ$     & bat             \\
                          & 2.02                          & 1.0391        & $4^\circ \times 5^\circ$     & bat             \\
                          & 2.12                          & 1.0021        & $4^\circ \times 5^\circ$     & bat             \\
                          & 2.17                          & 0.9848        & $4^\circ \times 5^\circ$     & (crash)         \\\hline
\multirow{5}{*}{10000}    & 1.0                           & 3.4389        & $4^\circ \times 5^\circ$     & rapid           \\
                          & 1.929                         & 2.1009        & $4^\circ \times 5^\circ$     & rapid           \\
                          & 2.12                          & 1.9573        & $4^\circ \times 5^\circ$     & rapid           \\
                          & 2.17                          & 1.9234        & $4^\circ \times 5^\circ$     & rapid           \\
                          & 2.22                          & 1.8908        & $4^\circ \times 5^\circ$     & rapid           \\
                          & 2.27                          & 1.8595        & $4^\circ \times 5^\circ$     & (crash)  
                          \\\hline 
\end{tabular}
\caption{Configuration of GCM simulations in this work.}
\label{table:configuration}
\end{table}

\begin{table}
\centering
\setlength\tabcolsep{4pt}
\begin{tabular}{ccccccc}
\hline \hline 
$T_{\mathrm{WD}}$ [K] & $F_{\mathrm{p}}$/$F_{\oplus}$ & $T_{s}${[}K{]} & TOA Net SW {[}$Wm^{-2}${]} & TOA Net LW {[}$Wm^{-2}${]} & Planetary Albedo & Rotation Regime \\\hline
\multirow{3}{*}{3500}             & 1.0                           & 259.60         & 206.85                     & 209.05                     & 0.3918           & bat             \\
                                  & 1.2                           & 308.60         & 281.76                     & 280.35                     & 0.3096           & -               \\
                                  & 1.3                           & -              & -                          & -                          & -                & (crash)         \\\hline
\multirow{5}{*}{5000}             & 1.0                           & 252.03         & 194.66                     & 197.00                     & 0.4273           & bat             \\
                                  & 1.2                           & 273.34         & 232.01                     & 233.36                     & 0.4311           & bat             \\
                                  & 1.3                           & 288.51         & 259.16                     & 258.93                     & 0.4135           & bat             \\
                                  & 1.35                          & 309.36         & 292.96                     & 290.01                     & 0.3615           & bat             \\
                                  & 1.4                           & -              & -                          & -                          & -                & (crash)         \\\hline
\multirow{5}{*}{6500}             & 1.0                           & 241.49         & 173.21                     & 176.84                     & 0.4904           & -               \\
                                  & 1.417                         & 278.39         & 250.47                     & 251.40                     & 0.4799           & bat             \\
                                  & 1.5                           & 288.86         & 269.66                     & 269.35                     & 0.4711           & bat             \\
                                  & 1.621                         & 309.21         & 302.83                     & 302.12                     & 0.4504           & bat             \\
                                  & 1.7                           & -              & -                          & -                          & -                & (crash)         \\\hline
\multirow{9}{*}{8000}             & 1.0                           & 216.47         & 127.40                     & 130.92                     & 0.6251           & rapid           \\
                                  & 1.417                         & 262.10         & 216.71                     & 220.13                     & 0.5499           & rapid           \\
                                  & 1.519                         & 270.41         & 233.28                     & 235.30                     & 0.5481           & rapid           \\
                                  & 1.621                         & 278.52         & 248.08                     & 249.57                     & 0.5496           & rapid           \\
                                  & 1.723                         & 284.14         & 258.21                     & 259.74                     & 0.5590           & rapid           \\
                                  & 1.965                         & 299.26         & 289.74                     & 288.37                     & 0.5661           & bat             \\
                                  & 2.02                          & 308.85         & 305.72                     & 304.82                     & 0.5546           & bat             \\
                                  & 2.12                          & 316.33         & 305.27                     & 305.31                     & 0.5762           & bat             \\
                                  & 2.17                          & -              & -                          & -                          & -                & (crash)         \\\hline
\multirow{5}{*}{10000}            & 1.0                           & 212.21         & 119.46                     & 123.55            & 0.6484           & rapid           \\
                                  & 1.929                         & 275.05         & 250.25                     & 252.22                     & 0.6182           & rapid           \\
                                  & 2.12                          & 295.92         & 289.06                     & 288.59                     & 0.5987           & rapid           \\
                                  & 2.17                          & 306.59         & 303.29                     & 300.37                     & 0.5886           & rapid           \\
                                  & 2.22                          & 333.34         & 295.43                     & 296.47                     & 0.6080           & rapid           \\
                                  & 2.27                          & -              & -                          & -                          & -                & (crash)
                                  \\\hline
\end{tabular}
\caption{Key output quantities for GCM simulations in this work. Quantities are computed based on 10-year averages after simulations reach statistical equilibrium.}
\label{table:quantity}
\end{table}

\clearpage

\subsection{Runaway Greenhouse Limit}
\begin{figure}[htbp]
    \centering
    \includegraphics[width=0.65\textwidth]{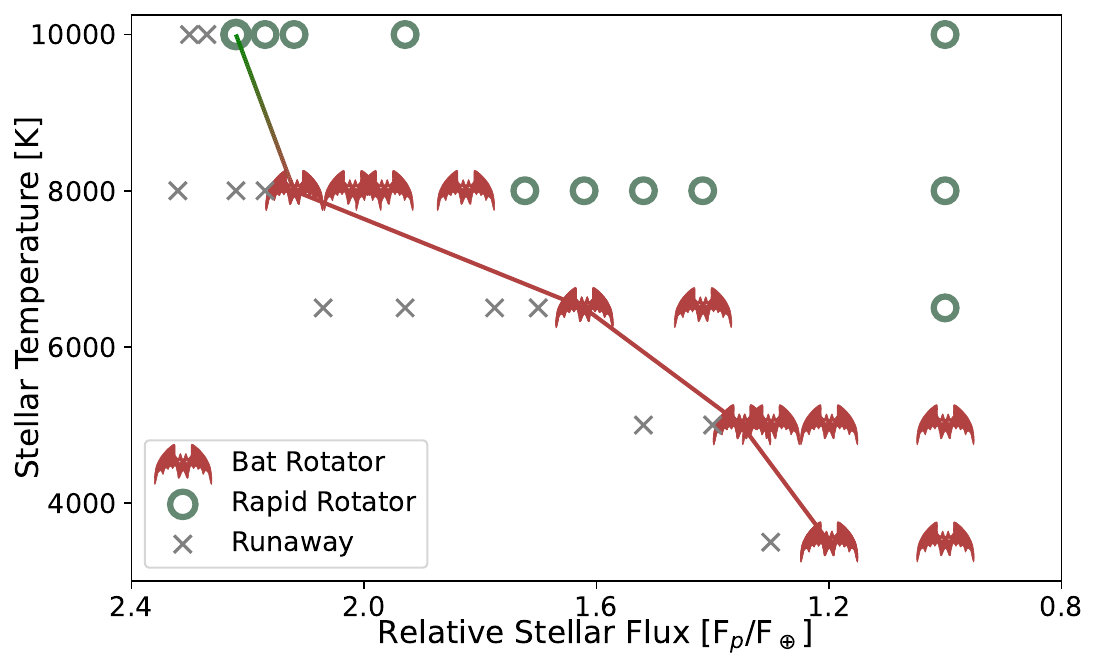}
    \caption{Runaway Greenhouse Limit around white dwarfs. The bat patterns and green circles represents the bat rotators and the rapid rotators which are converged simulations. The grey crosses are runaway simulations in our result.}
    \label{appenRGHL}
\end{figure}

Figure \ref{appenRGHL} and Table \ref{table:configuration} show how we determine the Runaway Greenhouse Limit (RGHL). At each stellar temperature, the RGHL is equal to the insolation of the last simulation that still converges (second-to-last row in each $T_\mathrm{WD}$ block in Table \ref{table:configuration}). Crosses in Figure \ref{appenRGHL} show other simulations that also crashed.

\newpage
\section{Appendix: Computing Thermal Flux}
\label{thermalflux}
Here we rederive the formula for 
the observer-projected flux stated in Appendix C of \citet{koll_deciphering_2015}.
Our derivation corrects a factor of $R^2/d^2$ which is missing from the expression in \citet{koll_deciphering_2015}; also note that the formula in \citet{koll_deciphering_2015} is not area-averaged, as it would otherwise be missing a factor of $\pi$. We additionally clarify the difference between flux and intensity: flux quantifies the amount of energy passing through a unit area per unit time, while intensity specifies the energy per unit area, time, and solid angle in a given direction. Our derivation below yields the same result as that obtained by \citet{seager_exoplanet_2010}.

\begin{figure}[t]
    \centering
    \includegraphics[width=0.4\textwidth]{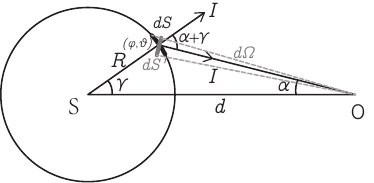}
    \caption{Sketch of a small area on a spherical planet, projected onto the observer's plane. dS: area element on the planet, located at ($\phi,\theta$). dS': projection of dS towards observer. Point O and S represent the observer and the center of the spherical planet.}
    \label{appen:compute_thermal}
\end{figure}

We assume the planet emits with isotropic intensity $I$. For any point ($\phi,\theta$) on the surface of the planet, a plane can be identified that passes through ($\phi,\theta$), the center of the planet, and the observer. We then project onto this plane, as shown in Figure \ref{appen:compute_thermal}. The angles $\gamma$ and $\alpha$ are related to $\phi$ and $\theta$ via
\begin{equation}
    \cos\gamma=\cos\phi\cos\theta,
\end{equation}
\begin{equation}
    \sin\alpha=R\frac{\sin(\alpha+\gamma)}{d}.
\end{equation}

Next we consider the solid angle $d\Omega$, which denotes how large a surface area on the sphere at ($\phi,\theta$), $dS=R^2\cos\theta d\theta d\phi$, appears to the observer. The projection of the surface area $dS$ in the direction of the solid angle is $dS'$, and the angle between $dS$ and $dS'$ is $\alpha+\gamma$, so
\begin{equation}
    dS'=dS\cos(\alpha+\gamma)=R^2\cos\theta d\theta d\phi\cos(\alpha+\gamma).
\end{equation}
According to the Law of Sines, the solid angle can be written as
\begin{equation}
    d\Omega=\frac{dS'}{d^2}\left( \frac{\sin(\alpha+\gamma)}{\sin\gamma} \right)^2.
\end{equation}
Here $\alpha$ is the angle between $d\Omega$ and direction of the observer. Assuming $d\gg R$ and $\lim \alpha\to 0$, we get $\cos\alpha\to 1$, $\sin(\alpha+\gamma)=\sin\alpha\cos\gamma+\cos\alpha\sin\gamma\to\sin\gamma$ and $\cos(\alpha+\gamma)=\cos\alpha\cos\gamma-\sin\alpha\sin\gamma\to\cos\gamma$. To derive the net flux received at distance $d$ by the observer, we integrate over the hemisphere facing the observer,
\begin{equation}
\begin{split}
    F_{\mathrm{obs}}&=\int I \cos\alpha d\Omega \\
    &=\int_{-\frac{\pi}{2}}^{\frac{\pi}{2}}\int_{-\frac{\pi}{2}}^{\frac{\pi}{2}} I(\phi,\theta) \cdot1\cdot\frac{\cos\gamma R^2\cos\theta d\theta d\phi}{d^2} \\
    &=\frac{R^2}{d^2}\int_{-\frac{\pi}{2}}^{\frac{\pi}{2}}\int_{-\frac{\pi}{2}}^{\frac{\pi}{2}} I(\phi,\theta) \cos\phi\cos^2\theta d\theta d\phi\\
    &=\frac{\pi R^2}{d^2}\frac{\int_{-\frac{\pi}{2}}^{\frac{\pi}{2}}d\theta\int_{-\frac{\pi}{2}}^{\frac{\pi}{2}}I(\phi,\theta)\cos\phi\cos^{2}\theta d\phi}{\int_{-\frac{\pi}{2}}^{\frac{\pi}{2}}d\theta\int_{-\frac{\pi}{2}}^{\frac{\pi}{2}}\cos\phi\cos^{2}\theta d\phi}.
\end{split}
\end{equation}

\bibliography{article}{}
\bibliographystyle{aasjournal}

\end{document}